\documentclass[11pt,a4paper]{article}
\pdfoutput=1
\usepackage{jheppub}
\usepackage[T1]{fontenc}
\usepackage{color}
\newcommand{\beq}{\begin{equation}}
\newcommand{\eeq}{\end{equation}}
\newcommand{\bea}{\begin{eqnarray}}
\newcommand{\eea}{\end{eqnarray}}
\newcommand{\dd}{\text{d}}
\newcommand{\bra}[1]{\Bigl\langle #1\Bigr|}
\newcommand{\ket}[1]{\Bigl|#1\Bigr\rangle}
\newcommand{\ldot}{\!\cdot\!}
\newcommand{\vep}{\varepsilon}
\newcommand{\ltap}{\;\raisebox{-.4ex}{\rlap{$\sim$}} \raisebox{.4ex}{$<$}\;}

\title{Coulomb gluons and the ordering variable}
\author{Ren\'e \'Angeles-Mart\'inez,}
\author{Jeffrey R. Forshaw}
\author{and Michael H. Seymour}
\affiliation{Consortium for Fundamental Physics, School of Physics \& Astronomy,\\
University of Manchester. Manchester M13 9PL. U.K.}

\emailAdd{rene.angelesmartinez@manchester.ac.uk}
\emailAdd{jeff.forshaw@manchester.ac.uk}
\emailAdd{michael.seymour@manchester.ac.uk}

\preprint{MAN/HEP/2015/17}
\abstract{We study in detail the exchange of a Coulomb (Glauber) gluon in the
  first few orders of QCD perturbation theory in order to shed
  light on their accounting to all orders. We find an elegant
  cancellation of graphs that imposes a precise ordering on
  the transverse momentum of the exchanged Coulomb gluon.  
}

\begin{document}

\maketitle
\flushbottom

\section{Introduction}
Large corrections to fixed-order matrix element calculations occur
in perturbative QCD as a result of soft and/or collinear parton
emissions. These can be calculated directly or using Monte Carlo event generators. The
latter are multi-purpose and capture some, but not all, of the leading
logarithmic behaviour via parton or dipole shower
evolution. Interference between
wide-angle soft gluon contributions can be included in the event
generator approach but at the expense of ignoring contributions that
are suppressed by powers of $1/N_c$. Most notably, Coulomb
(a.k.a.\ Glauber) gluon exchanges are ignored. 

In this paper we wish to study the physics of soft gluons beyond the
leading colour approximation, and of particular interest will be the correct
inclusion of Coulomb exchanges. It is well known that
Coulomb exchanges are ultimately responsible for diffractive processes
and the ambient particle production known as the ``underlying
event'' \cite{Gaunt:2014ska}. Moreover, attention has focussed on them due to the realization that they are the origin of the
super-leading logarithms discovered in gaps-between-jets
observables \cite{Forshaw:2006fk,Forshaw:2008cq} and later realized to
affect almost all observables in hadron-hadron
collisions \cite{Banfi:2010xy}, as well as being the origin of a
breakdown of collinear factorization \cite{catani2011,Forshaw:2012bi} in
hadron-hadron collisions.

Coulomb exchange should therefore be an important ingredient
in any reasonably complete description of the partonic final state of
hadron-hadron collisions. However, the
inclusion of Coulomb exchanges in the standard shower algorithms is
complicated because they mix colour and are non-probabilistic.
Although there is a framework capable in principle of encompassing these
corrections, \cite{Nagy:2007ty}, the actual implementation of
it \cite{Nagy:2012bt} neglects them, as do other attempts to include
sub-leading colour into parton showers \cite{Platzer:2012np}.

It is not entirely clear how
Coulomb exchanges should be included in an all-orders summation of
soft gluon effects. The aim of this paper is to show how they can be
included via a $k_T$-ordered evolution algorithm. We do not prove the
correctness of the algorithm to all orders in perturbation theory but
rather to the first two non-trivial orders. We think it is likely that
the procedure generalizes to all orders.

The algorithm, for a general observable, is built from the set of
cross sections corresponding to exclusive $n$ gluon emission, $\{ \dd
\sigma_n \}$:
\bea \label{eq:master}
\dd \sigma_0 &=& \bra{M^{(0)}} {\mathbf{V}}^\dag_{0,Q} {\mathbf{V}}_{0,Q} \ket{M^{(0)}} ~ \dd \Pi_0
\nonumber \\
\dd \sigma_1 &=& \bra{M^{(0)}} {\mathbf{V}}^\dag_{q_{1T},Q}\mathbf{D}_{1\mu}^\dag {\mathbf{V}}^\dag_{0,q_{1T}} {\mathbf{V}}_{0,q_{1T}}
\mathbf{D}_1^\mu  {\mathbf{V}}_{q_{1T},Q} \ket{M^{(0)}} ~ \dd \Pi_0 \dd \Pi_1
\nonumber \\
\dd \sigma_2 &=& \bra{M^{(0)}} {\mathbf{V}}^\dag_{q_{1T},Q}\mathbf{D}_{1\mu}^\dag {\mathbf{V}}^\dag_{q_{2T},q_{1T}}
\mathbf{D}^\dag_{2 \nu}  {\mathbf{V}}^\dag_{0,q_{2T}} {\mathbf{V}}_{0,q_{2T}} \mathbf{D}_2^\nu {\mathbf{V}}_{q_{2T},q_{1T}}
\mathbf{D}_1^\mu  {\mathbf{V}}_{q_{1T},Q} \ket{M^{(0)}} ~ \dd \Pi_0 \dd \Pi_1 \dd \Pi_2 
\hspace*{-1em}
\nonumber \\
& \mathrm{ etc.} & 
\eea
To reveal the underlying simplicity of the structure we have used a
very compact notation, which we now explain. The fixed-order matrix
element is represented by a vector in colour and spin, denoted
$\ket{M^{(0)}}$ and $\dd \Pi_0$ is the corresponding phase-space. Virtual
gluon corrections are encoded in the Sudakov operator:
\beq
{\mathbf{V}}_{a,b} = \exp \left[ -\frac{2\alpha_s}{\pi} \int_{a}^{b} \frac{\dd
    k_T}{k_T} \sum_{i<j} (-\bold{T}_i \cdot \bold{T}_j)
  \frac{1}{2} \left\{ \int \frac{\dd y \,\dd \phi}{2\pi} \omega_{ij} -
    i\pi \; \Theta(ij=II~\mathrm{or}~FF)\right\} \right]~, \label{eq:sudakov}
\eeq
where
\beq
\omega_{ij} = \frac{1}{2} k_T^2 \frac{p_i \cdot p_j}{(p_i \cdot k)(p_j
  \cdot k)}
\eeq
and the $\Theta$ term is unity for the case where partons $i$ and $j$
are either both in the initial state or both in the final state, and
zero otherwise (this is the term corresponding to Coulomb exchange).
The crucial ingredient of Eq.~(\ref{eq:sudakov}) is the
  fact that the limits on the transverse momenta of the virtual
  exchanges, $k_T$, are the transverse momenta of the emitted
  gluons. The colour charge of parton $i$ is denoted
$\bold{T}_i$, and $k_T$, $y$ and $\phi$ are the transverse
momentum, rapidity and azimuth of the virtual gluon with momentum $k$
that is exchanged between partons $i$ and $j$. The operator $\mathbf{D}_i^\mu$
corresponds to the real emission of a gluon with transverse momentum $q_{Ti}$ and
the associated phase-space element (including a factor $\alpha_s$ for convenience)  is $\dd \Pi_i$:
\bea
\mathbf{D}_i^\mu &=& \sum_j \bold{T}_j \; \frac{1}{2} q_{Ti}
\frac{p_j^\mu}{p_j \cdot q_i} ~,
\nonumber \\
\dd \Pi_i &=& -\frac{2\alpha_s}{\pi} \frac{\dd q_{Ti}}{q_{Ti}} \frac{\dd y_i
  \dd \phi_i}{2\pi} ~.
\label{eq:PS}
\eea
A general cross section can then be written
\beq
\sigma = \sum_{n=0}^{\infty} \int \dd \sigma_n \, F_n~,
\eeq
where $\{ F_n \}$ are functions of the phase-space that define the
observable. Although we have written formulae that are appropriate for
soft gluon corrections, it is straightforward to extend them to
include collinear emission too: the Sudakov operator ${\mathbf{V}}$ picks up a
hard-collinear piece and the splitting operator $\mathbf{D}$ is modified.

Equation (\ref{eq:master}) is expressed as a chain of real emissions
ordered in transverse momentum with Sudakov operators expressing the
non-emission at intermediate scales. If we would ignore the Coulomb
exchange contribution to the Sudakov operator then this would be the end
of the story, in the sense that Eq.~(\ref{eq:master}) encodes
well-known physics.
Moreover, if one takes the leading $N_c$ approximation
then the colour evolution is diagonal and this drastically simplifies
matters, allowing the computation of observables using a
cross-section level shower algorithm, e.g.\ as is done in an event
generator.

However, Coulomb exchanges are virtual corrections that do not
correspond to a non-emission probability. In QED they
exponentiate to an irrelevant phase in the scattering amplitude but
this does not happen in the case of non-Abelian QCD. Since Coulomb gluons have
transverse momentum but no rapidity or azimuth, it would seem most natural
to include them as in Eq.~(\ref{eq:master}). Indeed this is
exactly what we assumed in \cite{Forshaw:2006fk,Forshaw:2008cq,Keates:2009dn}, to compute the coefficient of the
coherence violating super-leading logarithmic term in the ``gaps
between jets'' observable. However, as pointed out in section 3.3 of \cite{Banfi:2010xy}, it is possible to
change the coefficient of the super-leading logarithm by limiting the
$k_T$ integral of the Coulomb exchange by some other function of the
real emission momenta. For example, the coefficient is divergent for
energy ordering, zero for angular ordering and one-half of the
$k_T$-ordered result in the case of virtuality ordering. 

In the remainder of this paper, we will demonstrate that
Eq.~(\ref{eq:master}) is correct, at least to the first few orders in
perturbation theory. To this end we will compute
the amplitudes for one and two real gluon emissions to
one-loop accuracy. Specifically, we perform Feynman-gauge
calculations in order to check the correctness of the operators
\beq
{\mathbf{V}}_{0,q_{1T}} \mathbf{D}_1^\mu  {\mathbf{V}}_{q_{1T},Q}  ~~~\mathrm{and}~~~{\mathbf{V}}_{0,q_{2T}} \mathbf{D}_2^\nu {\mathbf{V}}_{q_{2T},q_{1T}}
\mathbf{D}_1^\mu  {\mathbf{V}}_{q_{1T},Q}~. \label{eq:twoterms}
\eeq 
Since these expressions are to capture the leading soft
behaviour, we work within the eikonal approximation for emissions off
the fast partons involved in the hard sub-process. This is the only approximation we
make and, in particular, we use the full triple-gluon vertex for soft
gluon emissions off other soft gluons and we use the exact expressions
for soft-gluon propagators.  This means that we make no assumptions
about the relative sizes
of the momenta of real and virtual soft radiation. 

For simplicity, we focus mainly on the case where $\ket{M^{(0)}}$ corresponds
to two coloured incoming particles scattering into any number of colourless particles
(e.g.\ the Drell-Yan process)
and we only calculate the imaginary part of the loop integrals, since this corresponds to the
contribution from Coulomb gluon exchange. Of course Coulomb exchange
between the incoming hard partons is
irrelevant at the cross-section level for scattering into a colourless
final state, but our interest is at the amplitude level, where there
remains much to learn. In particular, our calculations are sufficient to reveal
the non-trivial way in which the real gluon transverse momenta serve
to cut off the Coulomb gluon momentum. Moreover, since we will keep the
full dependence on the colour matrices of the two incoming partons, our
results give a clear indication of the structure of the more
phenomenologically-interesting case of two coloured partons producing a
system of coloured partons. We will
perform the loop integrals over $k_T$ exactly and show that they result
in precisely the two ($k_T$-ordered) terms in (\ref{eq:twoterms}), up to
non-logarithmic corrections. We will also see how the non-Abelian nature of QCD plays a crucial role in
engineering the $k_T$ ordering.

Our focus in Section \ref{sec:one} is to make a check of the first term in
(\ref{eq:twoterms}), i.e.\ we consider the case of one real emission at
one loop order. This section concludes by pointing out that $k_T$ ordering does not arise from the simplicity
of the Drell-Yan process that we considered. Then we  study the case of two
real emissions, which provides a check of the second term in~(\ref{eq:twoterms}).
Firstly, in Section  \ref{sec:limtree}, we describe the kinematic
regions of interest and the behaviour
of the tree-level amplitude. Then, in Section \ref{sec:two}, we
move to the one-loop case. 

\section{One real emission \label{sec:one}}

\begin{figure}[ht]%
\centering
\includegraphics[width=0.55\textwidth]{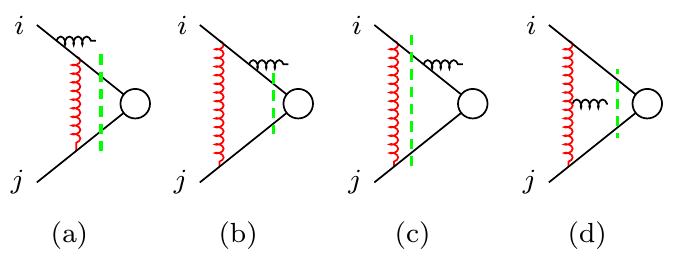}
\caption{Four cut graphs contributing to the amplitude for real
  emission of a gluon, with four-momentum $q_1$ and colour $c_1$, off fast
  parton $i$. There are
  three further graphs corresponding to emission off fast parton $j$.
\label{fig:one}}
\end{figure}

The imaginary part of the one-loop, one-emission amplitude can be
obtained from the cut graphs illustrated in Fig.~\ref{fig:one}. We
subsequently refer to cuts that pass through the two fast parton lines as
``eikonal cuts''. Note
that there are no contributions arising from cuts through a fast
parton and the Coulomb gluon, as we discuss briefly again towards the
end of this section.

The contribution to the amplitude from graphs (a)--(c) is
then\footnote{The normalization of the amplitude is consistent with
  the way we define the phase-space factor in Eq.~(\ref{eq:PS}). We
  use a colour-basis-independent notation \cite{Catani:1996vz,Forshaw:2008cq}
  for the colour state of the hard subprocess, but refer explicitly to
  the colour of the emitted gluon $c_1$, which means that we are not
  basis independent in the colour space of the soft gluons that dress
  the hard subprocess.}
\beq
-\frac{i \pi}{8\pi^2}  \frac{p_i \cdot \varepsilon_1}{p_i \cdot q_1}
\Bigl[  \bold{T}^{c_1}_i (\bold{T}_i \cdot \bold{T}_j)  - (\bold{T}_i \cdot
  \bold{T}_j)\bold{T}_i^{c_1}  + (\bold{T}_i \cdot \bold{T}_j) \bold{T}_i^{c_1}   \Bigr]  \int_0^{Q^2} \frac{\dd k_T^2}{k_T^2}
  \; \ket{M^{(0)}}~.
\eeq 
Although the contribution from graphs (b) and (c) cancels, it
is more instructive to keep them apart.

The notation is a little sloppy because we are not being clear
about the space in which the colour charge operators act, but it should
always be clear
from the context. The integral of the Coulomb gluon momentum, $k_T$, is
over the full range from 0 up to an ultraviolet scale that can be taken
to be the hard scale,~$Q$. Graph~(d) is
responsible for triggering the $k_T$ ordering. This is the only cut
graph involving the triple-gluon vertex and it gives rise to the contribution:
\beq
-\frac{i \pi}{8\pi^2} \frac{p_i \cdot \varepsilon_1}{p_i \cdot q_1}
\Bigl[ (\bold{T}_i \cdot \bold{T}_j)\bold{T}_i^{c_1} -\bold{T}_i^{c_1}
  (\bold{T}_i \cdot \bold{T}_j)     \Bigr] \int_0^{Q^2} \frac{\dd
  k_T^2}{k_T^2} \frac{q_{1T}^2}{k_T^2+q_{1T}^2} \;
   \ket{M^{(0)}}~.
\eeq 
Crucially, the
loop integral of graph~(d) acts as a switch. It is zero (i.e.\ sub-leading) if $k_T >
q_{1T}$ and when it is active it has the effect of exactly cancelling the
contribution from graphs (a) and (b). The result is that for $k_T >
q_{1T}$ only graph~(a) survives whilst for $k_T < q_{1T}$ only graph~(c)
survives, i.e.\ the final result is
\beq
-\frac{i \pi}{8\pi^2}  \frac{p_i \cdot \varepsilon_1}{p_i \cdot q_1}
 \left[ \bold{T}^{c_1}_i (\bold{T}_i \cdot \bold{T}_j)  \int_{q_{1T}^2}^{Q^2} \frac{\dd k_T^2}{k_T^2}
 +(\bold{T}_i \cdot \bold{T}_j) \bold{T}_i^{c_1} \int_0^{q_{1T}^2} \frac{\dd k_T^2}{k_T^2}\right]
     \ket{M^{(0)}}~.
\eeq 
These contributions, with the Coulomb gluon $k_T$
restricted to be bounded by the $q_{1T}$ of the real
emission are exactly in accordance with Eq.~(\ref{eq:master}), i.e.\ after
adding the contribution obtained after swapping $i \leftrightarrow j$
we get
\beq
 \Bigl[\bold{J}_1(q_1) \bold{C}_{q_{1T},Q} + \bold{C}_{0,q_{1T}}  \bold{J}_1(q_1)\Bigr]
\ket{M^{(0)}}~ , \label{eq:1ordering}
\eeq
where $\bold{J}_1(q_1)$ is the real emission operator:
\begin{align}
\bold{J}_1^{c_1}(q_1)\equiv \bold{T}^{c_1}_i \frac{ p_i \cdot \varepsilon_1}{p_i \cdot q_1}
+\bold{T}^{c_1}_j \frac{ p_j \cdot \varepsilon_1}{p_j \cdot q_1}, 
\end{align}
and the Coulomb exchange operator  $\bold{C}_{a,b}$ is
\begin{align}
\bold{C}_{a,b}\equiv -\frac{i\pi \,\bold{T}_i\cdot \bold{T}_j}{8\pi^2}  \int_{a^2}^{b^2} \frac{\dd k_T^2}{k_T^2}~.\label{eq:coulomb}
\end{align}
Of particular note is the way that the unwanted
cut of graph~(b) always cancels, either against graph~(c) or graph~(d).
Such a contribution would be problematic for any local evolution
algorithm, since it corresponds to a Coulomb
exchange retrospectively putting on-shell a pair of hard partons earlier in the
evolution chain. 

\begin{figure}[ht]%
\centering
\includegraphics[height=0.17\textwidth]{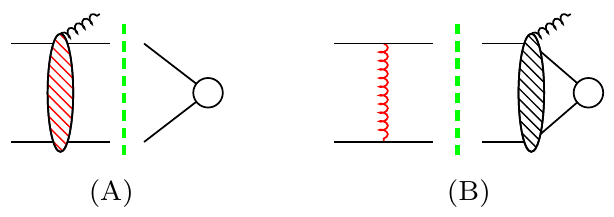}
\caption{The two cuts corresponding to the two different physical mechanisms
  for single gluon emission.
\label{fig:cut}}
\end{figure}

There is another way to think about this physics. The Coulomb exchange
corresponds to on-shell scattering of the incoming partons long before
the hard scattering and the real emission can occur either as part of
this initial-state scattering or, much later, as part of the
hard scattering. These two possibilities are illustrated in
Fig.~\ref{fig:cut}.

Graphs (a), (b) and (d) of Fig.~\ref{fig:one}  are of type~(A) and, in the domain where (d) is active, it 
cancels the other graphs. This means that the $k_T$ of
the Coulomb gluon must be greater than that of the real emission, i.e.\
it is as if the real emission is occurring coherently off a hard partonic
subprocess mediated by the Coulomb gluon. 
The sum over cuts of type~(A) gives
\beq
-\frac{i \pi}{8\pi^2} 
  \bold{T}_j^b (if_{c_1 ba})\bold{T}_i^a  \left[\frac{p_i \cdot \varepsilon_1}{p_i \cdot q_1}-\frac{p_j \cdot \varepsilon_1}{p_j \cdot q_1}\right]
 \int_{q_{1T}^2}^{Q^2} \frac{\dd k_T^2}{k_T^2}  \ket{M^{(0)}}~ 
 \label{eq:1type1}.
\eeq 
Graph~(c) is the only graph of type~(B). In this case the
real emission occurs much later than the Coulomb exchange, which therefore knows nothing of
the emission and so its $k_T$ can take any value, i.e.\
\beq
\bold{C}_{0,Q}\; \bold{J}_1(q_1) \ket{M^{(0)}}~.
 \label{eq:1type2}
\eeq
These contributions are separately gauge invariant, as can be seen by
making the replacement $\varepsilon_1 \to q_1$ in (\ref{eq:1type1}) and
noting also that $(\bold{T}_i + \bold{T}_j) \ket{M^{(0)}} = 0$
in (\ref{eq:1type2}), which is a statement of colour conservation.

There is a third type of cut that appears at intermediate steps of the
evaluation of some Feynman diagrams, in which the cut passes
through a fast parton and a soft gluon. This corresponds to an
unphysical (``wrongly time ordered'') process in which a gluon is
emitted during the hard process and this gluon scatters off one of the
incoming partons long before it was emitted. All such contributions to
each diagram cancel and we can neglect them, leaving only the cuts of
types~(A) and~(B).

To conclude this section, we comment on a trivial but important
generalization of expression (\ref{eq:1ordering}) to the case of
coloured particles
in the final state. Specifically, it follows that
\begin{align}
\Bigl[ \widetilde{\bold{J}}_1(q_1)\bold{C}_{q_{1T},Q} + \bold{C}_{0,q_{1T}}\widetilde{\bold{J}}_1(q_1) \Bigr] \ket{\widetilde{M}^{(0)}} \label{eq:g1ec}
\end{align}
is the imaginary part of the amplitude for one soft gluon emission off a general hard
sub-process with a Coulomb exchange in the initial state. Here, $\ket{\widetilde{M}^{(0)}}$ corresponds to two incoming hard
partons scattering into any number of hard coloured partons and
$\widetilde{\bold{J}}_1(q_1)$ is the total real emission operator,
i.e.\ including the cases where the gluon is emitted off a final-state
hard parton. This result  follows directly after noting that the
emission operator from final-state partons commutes with the Coulomb
exchange in the initial state and that  $\bold{C}_{0,Q} = 
\bold{C}_{0,q_{1T}}+\bold{C}_{q_{1T},Q}$.

\section{Two emissions at tree level \label{sec:limtree}}

We now turn to the case of two real emissions,  for which the transverse momentum ordering property is no longer an
exact result. Instead, it is a property of the amplitude in certain
regions of the phase-space of the emitted gluons. 
We will discuss
these regions in the next subsection, after which we will proceed to
study the behaviour of the amplitude at tree level. This will provide
the foundation for the calculation, which appears in the next section, of the
two-gluon emission amplitude with a Coulomb exchange.

\subsection{Phase-space limits \label{sec:limits}} 

Throughout this paper we will focus upon the following three limits. All of
them correspond to a strong ordering in the transverse momenta of the
real emissions, i.e. $q_{1T} \gg q_{2T}$.  In terms of light-cone variables\footnote{$q^{\pm}= (q_0\pm
q_1)/\sqrt{2}$ and $p_i=(p_i^+,0,\vec{0}_T)$,
$p_j=(0,p_j^-,\vec{0}_T)$.},  the three limits are:
\begin{itemize}
\item{Limit~1: Both emissions are at wide angle but one gluon is much softer
    than the other, i.e. $(q_1^\pm \sim q_{1T}) \gg (q_2^\pm \sim q_{2T})$.
  Specifically, we take $q_{2}\to \lambda q_2 $ and keep the leading
  term for small~$\lambda$. }
\item{Limit~2: One emission ($q_2$) collinear with $p_i$ by virtue of its small
    transverse momentum and the other ($q_1$) at a wide angle, i.e. $q_2^+
    \gg q_{2T} $ and   $q_1^+\sim q_{1T} \gg  q_{2T}$. Specifically,
    we take $q_2\to ( q_2^+ , \lambda^2 q_{2T}^2/(2q_2^+),
      \lambda \bold{q}_{2T})$ and keep the leading term for small~$\lambda$.}
\item{Limit~3: One emission ($q_1$) collinear with $p_i$ by virtue of its high
    energy and the other ($q_2$) at a wide angle, i.e. $q_1^+
    \gg q_{1T}$ and $q_{1T} \gg q_{2T} \sim q_2^+$. Specifically, we
    take\footnote{We use the eikonal
      approximation for the emitted gluons, in which the hard partons
      define light-like directions whose energies can be taken to be
      arbitrarily high. So even in the limit $\lambda\to0$, we assume
      $q_1^+/\lambda\ll p_i^+$.}
      $q_1 \to ( q_1^+/\lambda, \lambda q_{1T}^2/(2q_1^+),\bold{q}_{1T}) $ and
    $q_2 \to\lambda q_2$, and keep the leading
    term for small~$\lambda$.}
\end{itemize}
When we consider the leading behaviour of the amplitude, either at tree
or one-loop level, we
will make an expansion for small~$\lambda$,  keeping only the
leading terms. With the exception of Section \ref{sec:slowphysical},
we work with the following choice of polarisation vectors for the
emitted gluons:
\begin{eqnarray}
\varepsilon_\mu(q,\perp) &=&\frac{\epsilon_{\mu\nu\alpha\beta} q^\nu
  p_i^\alpha p_j^\beta}{\sqrt{2p_i\cdot p_j \, p_i\cdot q \, p_j\cdot q}}~,
                       \nonumber \\
\varepsilon^\mu(q,\parallel) &=& \frac{q\cdot p_j \, p_i^\mu -q\cdot
                                 p_i \, p_j^\mu- p_i\cdot p_j \, q^\mu  }
{\sqrt{2p_i\cdot p_j \, p_i\cdot q \, p_j\cdot q}}~.
\end{eqnarray}
In limits 2 and 3, only $\varepsilon^\mu(q,\parallel)$ of the collinear
parton, gives rise to a leading contribution. 

Limit~3 is of particular interest because it is the limit that gives
rise to the super-leading logarithms
\cite{Forshaw:2006fk,Forshaw:2008cq}.
It is worth noting that although $q_{1T} \gg q_{2T}$ in all three
limits, we may have $q_1^+ \sim q_2^+$ in limit~2 and $q_1^- \sim q_2^-$
in limit~3. This means that limits~2 and~3 are not sub-limits of
limit~1 in any trivial way. We will see that different Feynman diagrams contribute
differently in the different limits. It is therefore remarkable that the
final result is identical in all three limits. Although we have not
yet proven it, we suspect that the final results may well hold in the more general
case in which only  $q_{1T} \gg q_{2T}$.

\subsection{Tree-level amplitude \label{sec:tree2}}

\begin{figure}[ht]%
\centering
\includegraphics[height=0.2\textwidth]{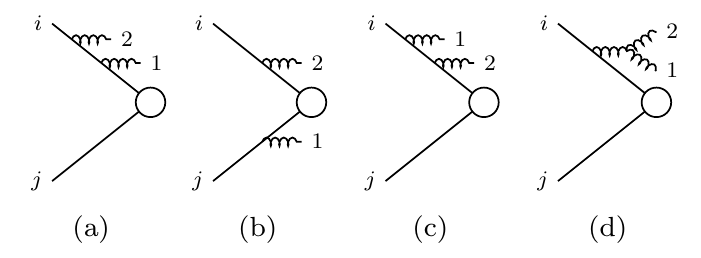} 
\caption{The case of two real emissions. There are four more graphs obtained by swapping $(i\leftrightarrow j)$.}\label{fig:apeI_graphs}
\end{figure}

The tree-level amplitude with two soft gluon emissions 
$\ket{M^{(0)}_2}$ can be expressed \cite{Catani:1999ss} in
terms of an operator $\bold{K}_2(q_1,q_2)$ that acts on the hard process to insert two real emissions,~i.e.
\begin{align}
\ket{M^{(0)}_2} = \bold{K}_2(q_1,q_2) \ket{M^{(0)}}~. 
\end{align}
In the case of only two incoming hard partons, we must consider the
four graphs shown in Fig.~\ref{fig:apeI_graphs} plus four further graphs
corresponding to the interchange $i \leftrightarrow j$.
As we will now show, $\bold{K}_2$ simplifies in each
of the limits 1--3 to a product of two single 
emission operators.

Let us consider first the leading behaviour in limit~1. In this 
region only graphs (a), (b) and (d) in Fig.~\ref{fig:apeI_graphs}
are leading. They give
\begin{align}
&\bold{K}^{c_1,c_2}_2(q_1,q_2) = \left[  \frac{ \bold{T}_i^{c_2} p_i
    \cdot\varepsilon_2}{ p_i \cdot q_2 } \frac{ \bold{T}_i^{c_1} p_i
    \cdot \varepsilon_1}{ p_i \cdot q_1 }  \right]
+ \left[  \frac{ \bold{T}_i^{c_2} p_i \cdot \varepsilon_2}{ p_i \cdot
    q_2 } \frac{ \bold{T}_j^{c_1} p_j  \cdot \varepsilon_1}{ p_j \cdot q_1 }   \right] \label{eq:ttreelim1}\\
&\hspace{5em}
+\left[    \frac{ if^{c_1c_2a}  q_1 \cdot \varepsilon_2 }{q_1 \cdot
    q_2}  \frac{ \bold{T}_i^{a} p_i \cdot \varepsilon_1}{ p_i \cdot q_1 } - 
\frac{if^{c_1c_2a} \bold{T}_i^{a} \varepsilon_1 \cdot \varepsilon_2
}{2q_1\cdot q_2} \right] + (i\leftrightarrow j)~.\nonumber 
\end{align}
The $\varepsilon_1\cdot \varepsilon_2$ term vanishes when it acts upon
 $\ket{M^{(0)}}$ due to colour conservation. The leading behaviour can
 thus be written
\begin{subequations}
\bea
\ket{M^{(0)}_2} &=& \bold{J}_{2}^{c_2c_1a}(q_2,q_1) \bold{J}_1^{a}(q_1) \ket{M^{(0)}}~,
\label{eq:treelim1} \\ 
\bold{J}_{2}^{c_2c_1a}(q_2,q_1) &\equiv&
\bold{J}_1^{c_2}(q_2)\delta^{c_1 a}+   \frac{if^{c_1c_2a}\; q_1\cdot
  \varepsilon_2 }{q_1\cdot q_2} \label{eq:J2} , 
\eea
\end{subequations}
where $\bold{J}_{2}^{c_2c_1a}(q_2,q_1)$ is the operator that adds a
second soft gluon $(q_2)$. 

In limit~2 only the first two graphs in Fig.~\ref{fig:apeI_graphs} 
are leading and they can be written
\begin{align}
&\ket{M^{(0)}_2} = \left[ \frac{
    \bold{T}_i^{c_2} p_i  \cdot \varepsilon_2}{ p_i\cdot q_2 } \right]
\bold{J}_{1}^{c_1}(q_1) \ket{M^{(0)}}~.\label{treelim2}
\end{align}
This is exactly what is obtained by taking the collinear limit
$q_2\parallel p_i$  in the expression for limit~1,
Eq.~\eqref{eq:treelim1}.

We now turn our attention to limit~3.
The leading contributions are graphs (a), (c) and (d), and the
$(i\leftrightarrow j)$ permutation of graph~(b) in
Fig.~\ref{fig:apeI_graphs}. These four contributions (in order) sum to
\begin{align}
&\bold{K}_2^{c_1,c_2}(q_1,q_2) = \left[   \frac{ \bold{T}_i^{c_2}  \varepsilon_2^-}{  q_2^- } \frac{ \bold{T}_i^{c_1} \varepsilon_1^-}{  q_1^-+q_2^- } \right] 
+\left[   \frac{ \bold{T}_i^{c_1}   \varepsilon_1^-}{  q_1^-} \frac{ \bold{T}_i^{c_2}   \varepsilon_2^-}{ q_1^-+q_2^- } \right]  \nonumber \\
&\hspace{5em}
+\left[    \frac{ if^{c_1c_2a}\;   \varepsilon_2^- }{ q_2^-}  \frac{ \bold{T}_i^{a}   \varepsilon_1^-}{ q_1^-+q_2^- }  \right]
+\left[\frac{ \bold{T}_j^{c_2}   \varepsilon_2^+}{  q_2^+ } \frac{ \bold{T}_i^{c_1}   \varepsilon_1^-}{ q_1^- }\right]~. \label{eq:treelim3}
\end{align}
At first glance it seems like an interpretation in terms of a product
of single emission operators is not possible any more.  However, using
$\bold{T}_i^{c_1}\bold{T}_i^{c_2}= \bold{T}_i^{c_2}\bold{T}_i^{c_1}
+if^{c_1c_2a} \bold{T}_i^a$,  the contribution of graph~(c) can
be written
\begin{align}
  \frac{ \bold{T}_i^{c_1}  \varepsilon_1^-}{  q_1^-} \frac{ \bold{T}_i^{c_2}   \varepsilon_2^-}{ q_1^-+q_2^- }
= \left[   \frac{ \bold{T}_i^{c_2}   \varepsilon_2^- }{ q_1^- }     \frac{ \bold{T}_i^{c_1}  \varepsilon_1^-}{ q_1^-+q_2^- } \right]
+\left[   \frac{ if^{c_1c_2a} \varepsilon_2^-}{  q_1^-} \frac{\bold{T}_i^a \varepsilon_1^-}{ q_1^-+q_2^- } \right]~.
\end{align}
The light cone variables make clear the fact that the two terms on the
right-hand side have the same dependence on colour and spin as the
first term on each line of Eq.~\eqref{eq:treelim3}.
Their momentum dependence can be combined using
\begin{align}
  \frac1{ q_1^- }\frac1{ q_1^-+q_2^- }+
  \frac1{ q_2^- }\frac1{ q_1^-+q_2^- }=
  \frac1{ q_1^- }\frac1{ q_2^- }~,
\end{align}
to give
\begin{align}
&\ket{M^{(0)}_2} = \bold{J}_{2\; }^{c_1c_2a}(q_2,q_1) 
\left(\frac{\bold{T}_i^{a} p_i  \cdot \varepsilon_1}{ p_i \cdot q_1}\right)
 \ket{M^{(0)}}~.\label{treelim3}
\end{align}
As in the case of limit~2, this can be obtained by taking the collinear
limit $q_1\parallel p_i$  in Eq.~\eqref{eq:treelim1}. Remarkably, we
will have the same property at one-loop order, i.e.\ the leading
expressions in limits 2 and 3 can be reached by taking the relevant
collinear limit of the leading expression in limit~1.
This is particularly non-trivial in limit~3, because the
leading graphs are not a subset of those in limit~1.

\begin{figure}[ht]%
\centering
\includegraphics[height=0.4\textwidth]{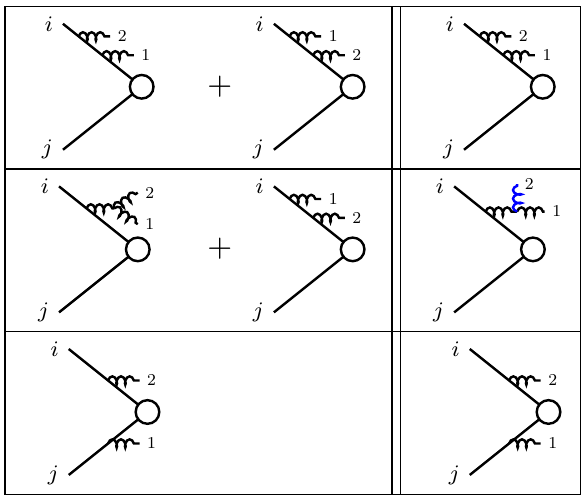} 
\caption{Diagrammatic representation of how to group the graphs in
  order to write the final result as a product of single emission
  operators. There are three further structures, obtained by exchanging
  $i\leftrightarrow j$.}
\label{fig:apeI_ordering}
\end{figure}

Figure~\ref{fig:apeI_ordering} shows how the graphs in
Fig.~\ref{fig:apeI_graphs} can be projected onto
three spin and colour structures. These particular structures are
special because the net projection onto 
each can be represented in terms of a product of two single emission
operators. 
Each grouping of graphs is associated with a specific spin and colour
structure, which can be read off from the graph at the end of
each row. These are
\begin{align}
 \left\{ \frac{\bold{T}_i^{c_2} p_i\cdot \varepsilon_2}{p_i \cdot q_2}\frac{\bold{T}_i^{c_1} p_i\cdot \varepsilon_1}{p_i\cdot q_1}  ,
\frac{if^{c_1c_2a}\; q_1\cdot \varepsilon_2}{q_1\cdot q_2} \frac{
  \bold{T}_i^{a} p_i\cdot \varepsilon_1}{ p_i \cdot q_1},  \frac{
  \bold{T}_i^{c_2} p_i \cdot  \varepsilon_2}{p_i \cdot q_2} \frac{
  \bold{T}_j^{c_1} p_j\cdot  \varepsilon_1}{p_j\cdot q_1}\right\} + \{(i\leftrightarrow j)\}~.
\end{align}
In limit~3, the two diagrams on each of the first two lines of
Fig.~\ref{fig:apeI_ordering} combine to give each effective diagram on
the right, interpreted \emph{as if\/} the two emissions were
independent. Equivalently, they conspire to act \emph{as if\/} $q_1^-$
and $q_2^-$ were strongly ordered, even though they are not. It is this
fact that allows the limit~3 result to be obtained from the limit~1
result (in which they are strongly ordered).

\section{Two emissions at one loop \label{sec:two}} 

We now consider the one-loop amplitude for a hard process with two
incoming partons and two soft emissions. 
In contrast to the single real emission case, we must now consider
graphs with cuts through two soft gluon lines, i.e.\ corresponding to a
Coulomb exchange between the two outgoing soft gluons.

\subsection{Eikonal cuts \label{sec:fast}}

\begin{figure}[ht]%
\centering
\includegraphics[height=0.17\textwidth]{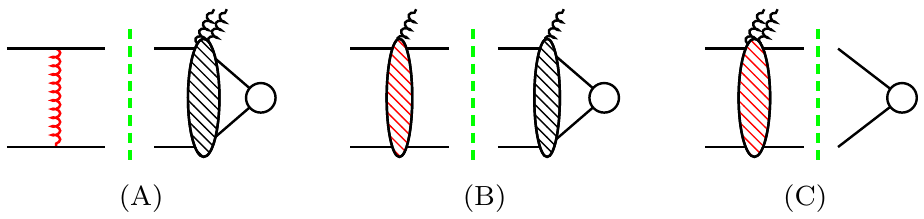}
\caption{The three cuts corresponding to the three different physical mechanisms
  for double-gluon emission. Each of these cuts is gauge invariant.
\label{fig:two}}
\end{figure}

Figure~\ref{fig:two} illustrates the three gauge-invariant classes of
cut graph, where the cut is through the two hard partons. As before, we refer to
these as eikonal cuts. The
corresponding amplitudes can be
reduced to transverse momentum integrals. In order to regulate the
diagrams that do not involve any emissions off the virtual gluon, we
introduce an ultraviolet cutoff of $Q^2$.
In all cases we regularize the
infrared divergences by analytically continuing the dimension of the
transverse momentum integral $\dd^{2}k_T\to \dd^{d-2} k_T$.  

\begin{figure}[ht]%
\centering
\includegraphics[height=1\textwidth]{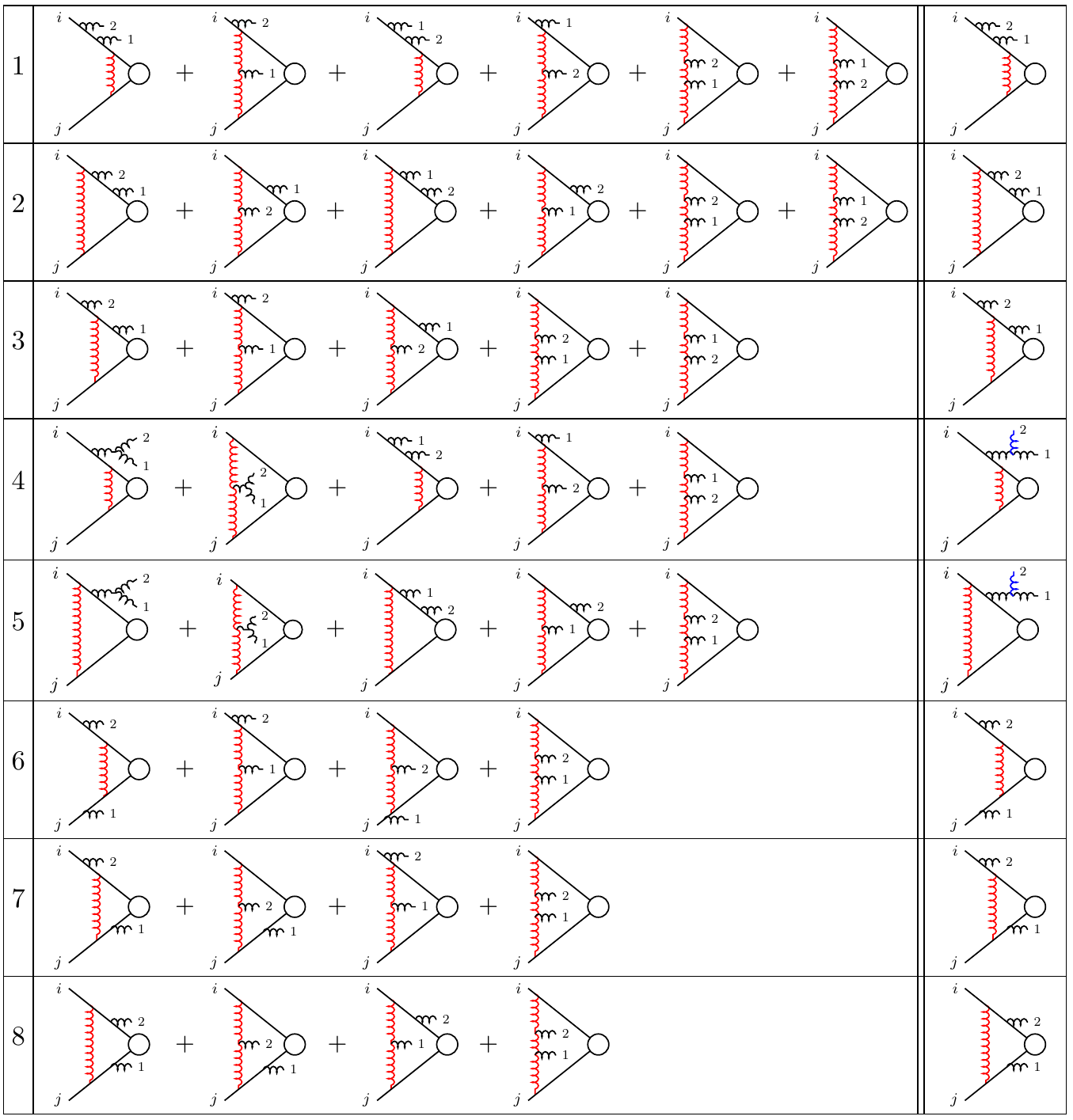} 
\caption{Diagrammatic representation of how to group the graphs that
  give rise to the transverse momentum-ordered expression in the case
  of two emissions at one loop. There are 12 more structures to consider: 8 are obtained by permuting 
  $(i\leftrightarrow j)$ and the other four are obtained by permuting
  $(1\leftrightarrow 2)$  and $(i\leftrightarrow j ,1\leftrightarrow 2 )$  in groups 3 and 7.}
\label{fig:master}
\end{figure}

We start by simply stating the bottom line. The remainder of this
section will be devoted to examining how these results arise. The
complete calculation involves explicitly computing the diagrams 
Fig.~\ref{fig:master} in dimensional regularization (using
\cite{Ellis:2011cr,Huber:2007dx}) and without any approximation except the eikonal
approximation.

The leading behaviour arising from eikonal cuts in limit~1 is
\begin{align}
 &\bigg[
\bold{C}_{0,q_{2T}} \bold{J}_{2}^{c_2c_1a}(q_2,q_1) \bold{J}_1^{a}(q_1)
+ \bold{J}_{2}^{c_2c_1 a}(q_2,q_1)\bold{C}_{q_{2T},q_{1T}}  \bold{J}_1^{a}(q_1)\nonumber\\
&+ \bold{J}_{2}^{c_2c_1a}(q_2,q_1) \bold{J}_1^{a}(q_1) \bold{C}_{q_{1T},Q}\bigg] \ket{M^{(0)}}~, \label{eq:ordlim1}
\end{align}
where the current $ \bold{J}_{2}^{c_2c_1a}(q_2,q_1)$ is defined in
Eq.~\eqref{eq:J2}. This expression is  the expected generalization of
the one-emission case \eqref{eq:1ordering} and the key point is that
the $k_T$ of the Coulomb exchange is ordered with respect to the
real-emission transverse momenta.  For the first two terms, the vector
$\bold{J}_1^{a}(q_1) \ket{M^{(0)}}  $ acts as a hard
subprocess for the second gluon emission, i.e.\ as in
Eq.~\eqref{eq:g1ec}  with $q_{1T}$  playing the role of $Q$.

Similarly, in limit~2 the sum over eikonal cuts gives
\begin{align}
 &\bigg[
\bold{C}_{0,q_{2T}} \left[\frac{\bold{T}_i^{c_2}\; p_i \cdot \varepsilon_2 }{p_i \cdot q_2}\right] \bold{J}_1^{c_1}(q_1)
+\left[\frac{\bold{T}_i^{c_2} \; p_i \cdot \varepsilon_2 }{p_i \cdot q_2}\right]  \bold{C}_{q_{2T},q_{1T}}  \bold{J}_1^{c_1}(q_1)\nonumber\\
&+ \left[\frac{\bold{T}_i^{c_2} \; p_i \cdot \varepsilon_2 }{p_i \cdot q_2}\right]  \bold{J}_1^{c_1}(q_1) \bold{C}_{q_{1T},Q}\bigg] \ket{M^{(0)}}~, \label{eq:ordlim2}
\end{align}
whilst in limit $3$ the result is
\begin{align}
 &\bigg[
\bold{C}_{0,q_{2T}} \bold{J}_{2}^{c_2c_1a}(q_2,q_1) 
\left[\frac{\bold{T}_i^{a}\; p_i \cdot \varepsilon_1 }{p_i \cdot q_1}\right]
+ \bold{J}_{2}^{c_2c_1 a}(q_2,q_1)\bold{C}_{q_{2T},q_{1T}}  
\left[\frac{\bold{T}_i^{a}\; p_i \cdot \varepsilon_1 }{p_i \cdot q_1}\right]
\nonumber\\
&+ \bold{J}_{2}^{c_2c_1a}(q_2,q_1) 
\left[\frac{\bold{T}_i^{a} \;p_i \cdot \varepsilon_1 }{p_i \cdot q_1}\right] \bold{C}_{q_{1T},Q}\bigg] \ket{M^{(0)}}~.\label{eq:ordlim3}
\end{align}
As in the tree-level case, the leading
behaviour in limits 2 and 3 coincides with the expressions that result
from taking the relevant collinear limit of the leading expression
in limit~1. These results confirm the conjecture  that
Eq.~\eqref{eq:master}  correctly reproduces the sum over eikonal cuts,
although, as we will shortly see, the way that the $k_T$ ordering
establishes itself is rather involved. 

\begin{figure}[ht]%
\centering
\includegraphics[height=0.25\textwidth]{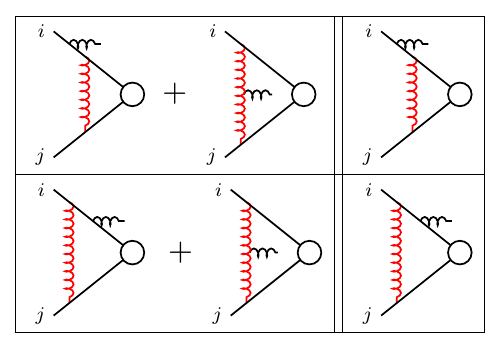} 
\caption{
Diagrammatic representation of how to group the graphs that
  give rise to the transverse momentum ordered expression 
  \eqref{eq:1ordering}. Two more structures are obtained by permuting
  $(i\leftrightarrow j)$.}\label{fig:master0}
\end{figure}

In order to see our way to Eq.~\eqref{eq:master} we must understand
how to deal with the graphs involving the triple-gluon vertex. In the
simpler case of only one real emission, this is illustrated in
Fig.~\ref{fig:master0}, which illustrates how the Feynman gauge
graphs are to be grouped together and projected onto the relevant spin
and colour tensors.

The corresponding amplitudes are
\begin{align}
\left\{  \frac{\bold{T}_i \;p_i \cdot \varepsilon }{ p_i \cdot q_1} \; \frac{-i\pi \; \bold{T}_j\cdot \bold{T}_i}{8\pi^2}
,   \frac{-i\pi \; \bold{T}_j\cdot \bold{T}_i}{8\pi^2}\;  \; \frac{\bold{T}_i \;p_i \cdot \varepsilon }{ p_i \cdot q_1}
\right\} + \left\{  (i\leftrightarrow j) \right\}.
\end{align}
The single graph involving the triple gluon vertex is thus shared out
between all four contributing tensors.

Figure~\ref{fig:master} is the generalization of Fig.~\ref{fig:apeI_ordering} and Fig.~\ref{fig:master0}. By way of
illustration, the tensor corresponding to the first graph in the first
row of the figure is
\beq
\; \frac{\bold{T}_i^{c_2} \; p_i \cdot \varepsilon_2}{p_i \cdot q_2} \;
 \frac{\bold{T}_i^{c_1} \;p_i \cdot \varepsilon_1}{ p_i \cdot q_1}  \;
 \frac{-i\pi \;\bold{T}_i \cdot \bold{T}_j }{8\pi^2} .\label{eq:line1}
\eeq 
In limits 1--3, every row in Fig.~\ref{fig:master} either adds up to
a subleading expression or to one of the terms in
Eqs.~\eqref{eq:ordlim1}--\eqref{eq:ordlim3}. This figure contains
all of the leading contributions arising from the 36 different
graphs with eikonal cuts.

In order to illustrate how the transverse momentum ordered integrals
arise, we will consider two examples in some detail.  We start by
taking a closer look at the first row of six graphs in
Fig.~\ref{fig:master}.
All of these graphs have only a single cut, corresponding to production
mechanism~(C) in
Fig.~\ref{fig:two}.  The first graph of the six gives rise to a factor 
of\footnote{In dimensional regularization, we write $\dd k_T^2 (k_T^2)^{-1} \to \mu^{2\epsilon} \dd k_T^2
  (k_T^2)^{-1-\epsilon} g(\epsilon)$, where
  $g(\epsilon)=1+\mathcal{O}(\epsilon)$.}
\beq
G_{11} =  \frac{q_1^-}{ ( q_{2}^-+q_{1}^-)}
  \int_0^{Q^2} \frac{\dd k_T^2}{k_T^2}~.
\eeq
 The factor multiplying the integral simplifies to unity in the case of limits 1 and 2 but not in
limit~3, where $q_1^-$ and $q_2^-$ could be of the same order. The projection of the third graph
gives
\beq
G_{13} =  \frac{q_{2}^-}{ (q_{1}^-+q_{2}^-)}
  \int_0^{Q^2} \frac{\dd k_T^2}{k_T^2}
\eeq
and this is only leading in the case of limit~3. Obviously these
Abelian-like contributions place no restriction on the $k_T$ of the
Coulomb exchange. Note that
\beq G_{11} + G_{13} =  \int_{0}^{Q^2} \frac{\dd
  k_T^2}{k_T^2}.\eeq
The second graph is the first involving the triple gluon vertex. It gives
\beq
G_{12} = - \left[
\int_0^{2 q_1^- q_2^+} \frac{\dd
  k_T^2}{k_T^2}
+ \frac{q_2^- - q_1^-}{q_2^- + q_1^-}  \int_0^{2(q_1^++q_2^-)^2 q_2^+/q_1^-} \frac{\dd
  k_T^2}{k_T^2}\right]~.
\eeq
We note that the Coulomb integral cannot be written purely in terms of
transverse momenta. However, the fourth graph is obtained from the
second by interchanging $q_1$ and~$q_2$.  Thus the sum
of graphs 2 and 4 is
\beq
G_{12} + G_{14} = -\frac{1}{(q_1^- + q_2^-)} \left[ 
q_2^- \int_{0}^{q_{2T}^2} \frac{\dd  k_T^2}{k_T^2} + q_1^-
\int_{0}^{q_{1T}^2} \frac{\dd  k_T^2}{k_T^2} \right]~. \label{eq:1314}
\eeq
Graphs 5 and 6 also combine to produce a reasonably compact result
involving the azimuthal angle between $q_{1T}$ and $q_{2T}$. It is sub-leading in limits 1 and 2, and in limit~3 it
simplifies to
 \beq
G_{15} + G_{16} \approx -\frac{q_2^-}{ (q_1^- + q_2^-)} \int_{q_{2T}^2}^{q_{1T}^2} \frac{\dd  k_T^2}{k_T^2}~.
\eeq 
Now we can combine the graphs. In limits 1 and 2 only $G_{11}$ and
$G_{14}$ contribute, with the latter contributing only for $k_T <
q_{1T}$, exactly as in the one emission case. The two combine to give 
\beq
 \int_{q_{1T}^2}^{Q^2} \frac{\dd  k_T^2}{k_T^2}~,
\label{eq:g1}
\eeq
which is exactly as expected. Limit~3 is more subtle and involves the
interplay of all 6 graphs. Remarkably, the sum of these is also
exactly equal to
(\ref{eq:g1}). The key is the way graphs 5 and 6 serve to extend
the upper limit in the first of the two terms in Eq.~(\ref{eq:1314})
from $q_{2T}^2$ to $q_{1T}^2$, so that the net effect of all four
graphs involving the triple-gluon vertex is merely to cut out the region with
$k_T < q_{1T}$.

It is also instructive to look at the graphs in the third row of Fig.~\ref{fig:master}. These 
involve cuts of type (B) and (C) in Fig.~\ref{fig:two}. We will just state the results (the
subscripts $B$ and $C$ refer to the cut):
\beq
G_{31B} =  \int_0^{Q^2} \frac{\dd k_T^2}{k_T^2}
= -G_{31C}~,
\eeq
\beq
G_{32} + G_{33C} = \frac{1}{(q_1^- + q_2^-)} \left[ 
q_2^- \int_{0}^{q_{2T}^2} \frac{\dd  k_T^2}{k_T^2} + q_1^-
\int_{0}^{q_{1T}^2} \frac{\dd  k_T^2}{k_T^2} \right],
\eeq
\beq
G_{33B} = - \int_0^{q_{2T}^2} \frac{\dd k_T^2}{k_T^2}~,
\eeq
so that
\beq
G_{32} + G_{33} =  \frac{q_1^-}{(q_1^-+q_2^-)} \int_{q_{2T}^2}^{q_{1T}^2} \frac{\dd k_T^2}{k_T^2}~.
\eeq
Once again the graphs where both gluons are emitted off the Coulomb
exchange are sub-leading in limits 1 and 2 and in limit~3 we find
\beq
G_{34}+G_{35} \approx  \frac{q_2^-}{(q_1^-+q_2^-)} \int_{q_{2T}^2}^{q_{1T}^2} \frac{\dd k_T^2}{k_T^2}~.
\eeq
On summing the graphs we obtain the expected result:
\beq
 \int_{q_{2T}^2}^{q_{1T}^2} \frac{\dd
  k_T^2}{k_T^2}.
\eeq
Notice how the sum of type (B) cuts is exactly as expected from the single-gluon
emission case, i.e.\ the Coulomb exchange satisfies $k_T > q_{2T}$.

\subsection{Physical picture \label{sec:2physical}}

As we did for the one-emission amplitude, it is interesting to group
together the cut graphs into gauge-invariant sets. In this case, that
means according to the cuts shown in Fig.~\ref{fig:two}. Cuts (A) and (B) are quite straightforward because
they can be deduced directly from the one real-emission case. In (A), the
Coulomb exchange occurs long before the double-emission and its $k_T$ is
unbounded (see Eq.~\eqref{eq:1type2}); the result (which is exact in
the eikonal approximation) is
\begin{align}
\bold{C}_{0,Q} \bold{K}^{c_1,c_2}(q_1,q_2) \ket{M^{(0)}}~, \label{eq:type1}
\end{align}
where $\bold{K}^{c_1,c_2}(q_1,q_2)$ is the double-emission operator, introduced in Section \ref{sec:tree2}. The gauge invariance of this expression is inherited from the gauge invariance of the 
tree-level double emission amplitude, $\bold{K}(q_1,q_2)\vert
M^{(0)}\rangle$.

In the case of cut (B), one of the emissions occurs
together with the Coulomb exchange, long before the hard scatter, and the
other during the hard scatter. These could be $q_{1,2}$ either way
round. In the case that it is $q_1$ that is emitted with the Coulomb
exchange, just like the case of cut (A) in Fig.~\ref{fig:cut}, its
$k_T$ must be larger than that of the real emission, $k_T >q_{1T}$ (see
Eq.~\eqref{eq:1type1}):
\beq
\left[ -\frac{i \pi}{8\pi^2} 
  \bold{T}_j^b if^{bc_1a}\bold{T}_i^a  \left[\frac{p_j \cdot \varepsilon_1}{p_j \cdot q_1}-\frac{p_i \cdot \varepsilon_1}{p_i \cdot q_1}\right]
 \int_{q_{1T}^2}^{Q^2} \frac{\dd k_T^2}{k_T^2} \right]  \bold{J}^{c_2}(q_2)
  \ket{M^{(0)}}~, \label{eq:type2}
\eeq 
which is manifestly gauge invariant. 

Cut (C) involves physics that cannot be inferred from the one-emission
amplitude. In view of Eq.~\eqref{eq:1type1}, one might anticipate that
this contribution is also infrared finite and this is indeed the
case. The proof of this involves the 
graph containing the four-gluon vertex, which is subleading in limits
1--3. The leading expression in limit~1 is
\begin{align}
-\frac{i\pi}{8\pi^2} &
\left[ 
\frac{p_j\cdot \varepsilon_1}{p_j\cdot q_1}-\frac{p_i\cdot \varepsilon_1}{p_i\cdot q_1}
\right]\bigg{\{}
\left[ 
\frac{p_j\cdot \varepsilon_2}{p_j\cdot q_2}-\frac{q_1\cdot \varepsilon_2}{q_1\cdot q_2}
\right] \left[ \bold{T}_{j}^{d}if^{dc_2b} if^{bc_1a}\bold{T}_i^a \right]\nonumber\\
&+\left[ 
\frac{q_1\cdot \varepsilon_2}{q_1\cdot q_2}-\frac{p_i\cdot \varepsilon_2}{p_i\cdot q_2}
\right] \left[ \bold{T}_{j}^{d}if^{dc_1b} if^{bc_2a}\bold{T}_i^a \right]\bigg\}
\int_{q_{1T}^2}^{Q^2} \frac{\dd k_T^2}{k_T^2} \ket{M^{(0)}}~.\label{eq:type3}
\end{align}
This is manifestly gauge invariant and, as anticipated, the result is
cut off from below by
 the larger of the two emitted transverse momenta. Note that
 Eq.~(\ref{eq:type3}) can be obtained directly
 by considering the coherent emission of $q_2$ off the $2 \to 3$ process described by  Eq.~(\ref{eq:1type1}).
As was the case at tree
level,  the leading behaviour of the expressions in limits  $2$ and
$3$ can be deduced by taking the respective collinear limits of this expression. By using the algebra of
the generators one can show  that the sum of Eq.~\eqref{eq:type1}, Eq.~\eqref{eq:type2} and its permutation $(1\leftrightarrow2 )$
and Eq.~\eqref{eq:type3} is equal to
\eqref{eq:ordlim1}, \eqref{eq:ordlim2} and \eqref{eq:ordlim3} in limits 1--3
respectively.

It is quite straightforward to generalize this entire section to include the case of a hard process with outgoing hard
partons and a Coulomb exchange between the two incoming hard
partons.\footnote{In order to confirm Eq.~\eqref{eq:master} for a
  general hard process, we need also to consider Coulomb exchanges in
  the final state. We leave this to future work.}  

\subsection{Soft gluon cuts\label{sec:slow}}

\begin{figure}%
\centering
\includegraphics[height=0.15\textwidth]{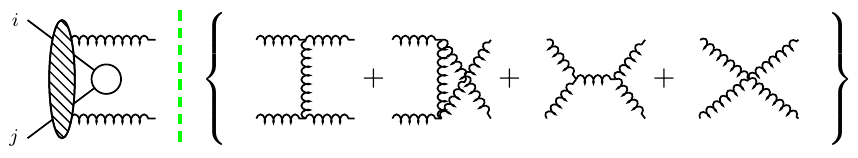}
\caption{Kinematically allowed soft gluon cuts. 
}\label{fig:gc}
\end{figure}

To complete the analysis, we turn our attention to the ``soft gluon
cuts'' illustrated in Fig.~\ref{fig:gc}. We will show
that the leading behaviour in the limits 1--3 is again in agreement
with Eq.~\eqref{eq:master}.

Before presenting the full result, it is useful to focus first
only on the $1/\epsilon$ poles.
\begin{figure}[ht]%
\centering
\includegraphics[height=0.15\textwidth]{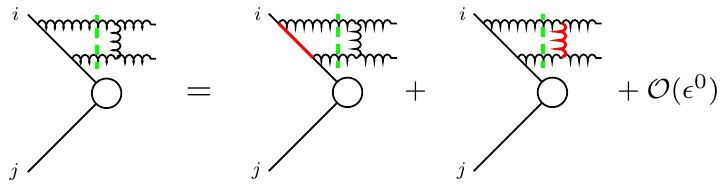}
\caption{Schematic representation of regions that give rise to
  infrared poles. Poles arise as a result of the vanishing of the red propagators.}
\label{fig:D6}
\end{figure}
In general the integrals of these cut graphs contain more than one
region in which the propagators vanish and, in dimensional 
regularization, each region gives rise to a $1/\epsilon$ pole. To
illustrate the point, we consider the first cut graph in
Fig.~\ref{fig:D6}, which gives
\begin{align}
 \frac{1}{2}\bold{T}_i^a  if^{ac_2b} if^{bc_1d} \bold{T}_i^{d} \int \frac{\dd^{d}l}{(2\pi)^d}  
\frac{if(l,p_i,q_1,q_2)\,(2\pi)\delta^{+} (l)
  \,(2\pi)\delta^{+}(q_1+q_2-l)}{(p_i\cdot l ) (l \cdot q_1 )}\,, \label{eq:D6}
\end{align}
where $f$ is a scalar function whose precise form is not important
and $\delta^+(l) = \theta(l_0)\delta(l^2)$. In the
reference frame in which the time-like vector $q_1+q_2$ is 
at rest, one can integrate over the energy $l_0$ and the magnitude 
of the $(d-1)$-momentum $\vert \vec{l} \vert$ to give
\begin{align}
  \frac{1}{2}\bold{T}_i^a  if^{ac_2b} if^{bc_1d} \bold{T}_i^{d}\,
  \frac{((q_1^0+q_2^0)/2)^{d-6}}{8 (2\pi)^{d-2} p_i^0 q_1^0} 
 \int \dd \Omega_{d-2} \frac{if(l)   }{ [1- \hat{p}_i\cdot\hat{l} ][ 1- \hat{q}_1\cdot\hat{l}] }\,,
\end{align}
where $\dd\Omega_{d-2}$ is the solid angle element of the unit
$(d-2)$-sphere. Clearly the denominator of the integrand only vanishes
when the virtual light-like momentum is either collinear with $p_i$ or
$q_1$, which cannot occur simultaneously\footnote{Unless 
these two vectors are exactly collinear, but we are excluding this case.}. It follows that the pole part of this expression 
can be computed as
\bea
 \int \dd \Omega_{d-2} \frac{f(l)  }{ [1- \hat{p}_i\cdot\hat{l} ][ 1- \hat{q}_1\cdot\hat{l}] }
 &=& \frac{f(l)\vert_{\hat{l}= \hat{p}_i} }{1-\hat{p}_i\cdot  \hat{q}_1}
 \int \frac{\dd\Omega_{d-2}}{[ 1- \hat{p}_i\cdot\hat{l}]} \nonumber \\ &+ &
 \frac{f(l)\vert_{\hat{l}= \hat{q}_1} }{1-\hat{p}_i\cdot  \hat{q}_1} \int \frac{\dd\Omega_{d-2}}{[ 1- \hat{q}_1\cdot\hat{l}]} +\mathcal{O}(\epsilon^0).
\eea
The remaining angular integration  can be performed by standard
methods, after which, Eq.~\eqref{eq:D6} can be written
\begin{align}
  \bold{T}_i^a  if^{ac_2b} if^{bc_1d} \bold{T}_i^{d}\,
  \frac{-i\pi}{8\pi^2} \frac{1}{(-p_i\cdot q_1) (q_1+q_2)^2} \left[
   -\frac{f(l)}{\epsilon}\Bigg{\vert}_{l= \frac{q_1\cdot
       q_2}{p_i\cdot (q_1+q_2)}p_i}-\frac{f(l)}{\epsilon}\Bigg{\vert}_{l=
     q_1} \right] +\mathcal{O}(\epsilon^0). 
  \label{eq:Dnext}
\end{align}
This expression indicates that the pole part of this cut graph arises
from the region in which the virtual emission is collinear to the hard
momentum $l^\mu\to \frac{q_1\cdot q_2}{p_i\cdot (q_1+q_2)}p_i^\mu$ 
and from the collinear region  $l^\mu\to q_1^\mu$. The latter corresponds to an infinitely soft 
virtual exchange between the two real emissions.  These two
contributions are represented on the right-hand side of
Fig.~\ref{fig:D6}.

Exactly the same type of analysis can be carried out to compute the
pole parts of each of the cut graphs in Fig.~\ref{fig:gc}. In all
cases, the $1/\epsilon$ poles arise either from the
region in which one of the eikonal propagators vanishes (collinear
singularities) or from the region in which the two real emissions
exchange a soft gluon between them. 
We note that included in Fig.~\ref{fig:gc} are cut self-energy
graphs and the corresponding ghost graphs should be added to these.
However, neither of them gives rise to infrared poles (or their
associated logarithms).
 
\begin{figure}[ht]%
\centering
\includegraphics[height=0.165\textwidth]{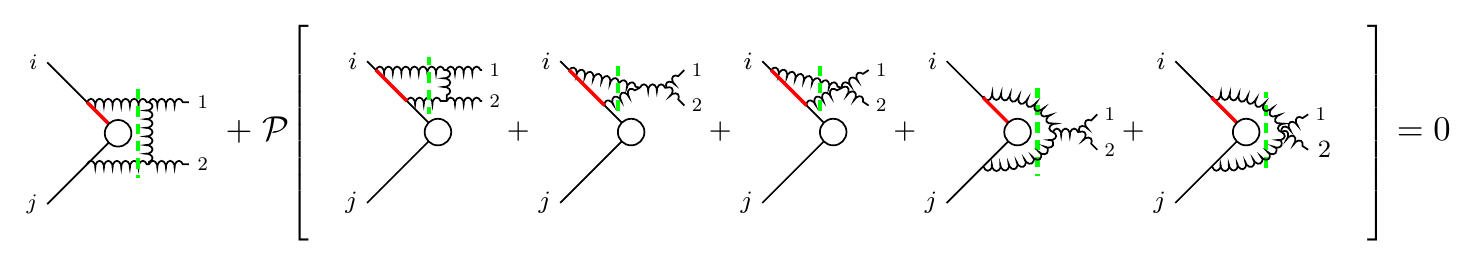}
\caption{Cancellation of collinear poles. The operator
$\mathcal{P}$ projects out the appropriate colour structure.}
\label{fig:polesij}
\end{figure}

\begin{figure}[ht]%
\centering
\includegraphics[height=0.15\textwidth]{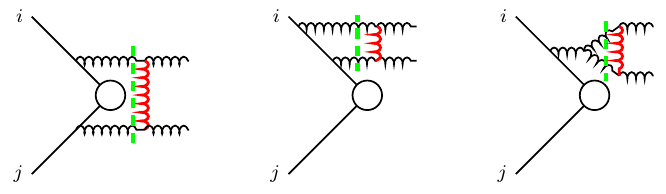}
\caption{The relevant $1/\epsilon$ poles arising from soft-gluon cuts
  correspond to Coulomb exchange between the two real emissions.}
\label{fig:polesh1}
\end{figure}

The colour operator
associated with each leading
graph in Fig.~\ref{fig:gc} can be written as a linear combination of
the colour structure on the left-hand side of
Fig.~\ref{fig:polesij} and its permutation $(1\leftrightarrow 2)$. For example, the colour operator
corresponding to the graph in Fig.~\ref{fig:D6} can be rewritten as
\begin{align}
\bold{T}_i^a  if^{ac_2b} if^{bc_1d} \bold{T}_i^{d} \ket{M^{(0)}}= - \bold{T}_j^d    if^{dc_1b} if^{bc_2a}  \bold{T}_i^{a}\ket{M^{(0)}}~.
\end{align}
After
expressing all of the colour structures in this way, one can confirm
that the poles corresponding to collinear
singularities
cancel. This cancellation gives rise to the zero on the right-hand side of Fig.~\ref{fig:polesij}. 
It follows that the only $1/\epsilon$ poles of the cut
graphs in Fig.~\ref{fig:gc} arise from a Coulomb exchange between
the two real emissions. These are represented in 
Fig.~\ref{fig:polesh1}. 
Explicitly, the pole part of the amplitude arising from the sum
over all soft gluon cuts can 
written
\beq
-\frac{1}{\epsilon} \left[  \frac{-i \pi}{ 8\pi^2 } if^{c_1 e a_1} if^{c_2 e a_2}  \right] \bold{K}_2^{a_1a_2}(q_1,q_2) \ket{M^{(0)}},
\label{eq:polesgc}
\eeq
where $\bold{K}_2(q_1,q_2) $ is the two-gluon emission tensor.
This expression can be combined with the pole part of 
Eq.~\eqref{eq:type1} to determine the leading $1/\epsilon$ pole of the
imaginary part of the double-emission amplitude. 

We will now go beyond the calculation of the leading $\epsilon$ poles
and compute the corresponding leading logarithmic contribution arising
from the soft gluon cuts. As before we will compute all of the
contributing Feynman graphs exactly in dimensional regularisation and
within the eikonal approximation, and then extract the leading
behaviour in limits 1--3. To do this we
make use of \cite{Ellis:2011cr,Huber:2007dx,vanNeerven}.
Recall that the colour part of all of
the graphs can be written as a linear combination of the colour
structure of the graph on the left-hand side of Fig.~\ref{fig:polesij}
and its permutation $(1\leftrightarrow 2)$. Written in terms of these
two colour tensors,
in limit~1, the amplitude is
\bea
&-&\frac{i \pi}{ 8\pi^2 }\left[ \frac{p_j\cdot \varepsilon_1}{p_j\cdot
    q_1}-\frac{p_i\cdot \varepsilon_1}{p_i\cdot q_1}\right]  \left\{
\left[ \frac{p_j\cdot \varepsilon_2}{p_j \cdot q_2}-\frac{q_1\cdot
    \varepsilon_2}{q_1 \cdot q_2}\right]
\left[-\frac{1}{\epsilon}+\ln\left(\frac{2 q_2\cdot q_1\, p_j\cdot q_2
    }{q_1\cdot p_j \; \mu^2}\right) \right]
\bold{T}_j^{d}if^{dc_2b}if^{bc_1 a}\bold{T}_i^{a} \right.
 \nonumber \\ 
&-& \left. \left[ \frac{p_i\cdot \varepsilon_2}{p_i \cdot q_2}-\frac{q_1\cdot
     \varepsilon_2}{q_1 \cdot q_2}\right]
\left[-\frac{1}{\epsilon}+\ln\left(\frac{2 q_2\cdot q_1 \, p_i\cdot q_2 }{q_1\cdot p_i\; \mu^2}\right)  \right] \bold{T}_j^{d}if^{dc_1b}if^{bc_2 a}\bold{T}_i^{a}
\right\} \ket{M^{0}}~. \label{eq:lim1gce}
\eea
This expression is manifestly gauge invariant but, at first sight, it
looks quite different from Eq.~\eqref{eq:polesgc}. As we will discuss
in more detail in the next section, each of the logarithmic terms can be
written in terms of the transverse momentum of gluon 2 measured in
either the $p_j\!+\!q_1$ or $p_i\!+\!q_1$ rest frame:
\beq
  q_{2T(i1)}^2 = \frac{2 q_2\cdot q_1 \, p_i\cdot q_2 }{q_1\cdot p_i}\,.
\eeq
In order to compare Eq.~(\ref{eq:lim1gce}) with Eq.~\eqref{eq:polesgc}, it is
convenient to introduce the rapidity:
\beq
q^{\pm}_i= q_{iT} e^{\pm y_i} /\sqrt{2}~.
\eeq
The logarithms are then
\begin{align}
\ln\left(\frac{2 q_2\cdot q_1\,p_i\cdot q_2 }{q_1\cdot p_i \mu^2}\right)
=\ln\left(\frac{ q_{2T(i1)}^2 }{\mu^2}\right)
=\ln\left(\frac{ q_{2T}^2 }{\mu^2}\right)+ 
\ln\left(\frac{2q_1\cdot q_2}{q_{1T} {q_{2T}}}  \right) +y_1-y_2,\\
\ln\left(\frac{2 q_2\cdot q_1\,p_j\cdot q_2 }{q_1\cdot p_j \mu^2}\right)
=\ln\left(\frac{ q_{2T(j1)}^2 }{\mu^2}\right)
=\ln\left(\frac{ q_{2T}^2 }{\mu^2}\right)+ 
\ln\left(\frac{2q_1\cdot q_2}{q_{1T} {q_{2T}}}  \right) +y_2-y_1,
\end{align}
and we see that the two are equal up to formally sub-leading terms ($\propto
y_i$). Eq.~\eqref{eq:lim1gce} can therefore be simplified and, using
colour conservation and the colour algebra, written as
\begin{align}
\left\{ -\frac{i \pi}{ 8\pi^2 } if^{c_1 e a_1} if^{c_2 e a_2}   
\left[ -\frac{1}{\epsilon} +\ln\left(\frac{ q_{2T}^2 }{\mu^2}\right) 
+\ln\left( \frac{2q_1\cdot q_2}{q_{1T} {q_{2T}}}  \right)\right]
 \right\} \; \bold{K}_2^{a_1a_2}(q_1,q_2)\ket{M^{(0)}}~.\label{eq:gclim1}
\end{align} 
The operator enclosed in curly brackets has the colour structure of a Coulomb exchange 
between the two soft real emissions, and its pole part agrees with Eq.~\eqref{eq:polesgc}.
The first logarithm can be written as 
\begin{align}
-\frac{1}{\epsilon} +\ln\left(\frac{ q_{2T}^2 }{\mu^2}\right) +\mathcal{O}(\epsilon)= 
\int_0^{q_{2T}^2} \frac{ \mu^{2\epsilon} \; \dd l_{T}^2}{(l_{T}^2)^{1+\epsilon}},
\end{align}
which is in agreement with Eq.~(\ref{eq:master}), and the second
logarithm is sub-leading. 

In limits 2 and 3, the sum over soft gluon cuts can be written as 
\beq
\left\{  -\frac{i \pi}{ 8\pi^2}  if^{c_1 e a_1} if^{c_2 e
    a_2}  \int^{q_{2T}^2}_0  \frac{ \mu^{2\epsilon} \; \dd l_{T}^2}{(l_{T}^2)^{1+\epsilon}}
 \right\}  \bold{K}_2^{a_1a_2}(q_1,q_2) \ket{M^{(0)}} \label{eq:gclim2}~.
\eeq
Once again, the result in
limits 2 and 3 can be deduced by taking the corresponding 
collinear limit of the leading expression in limit~1,
Eq.~\eqref{eq:lim1gce}.

\begin{figure}[ht]%
\centering
\includegraphics[height=0.45\textwidth]{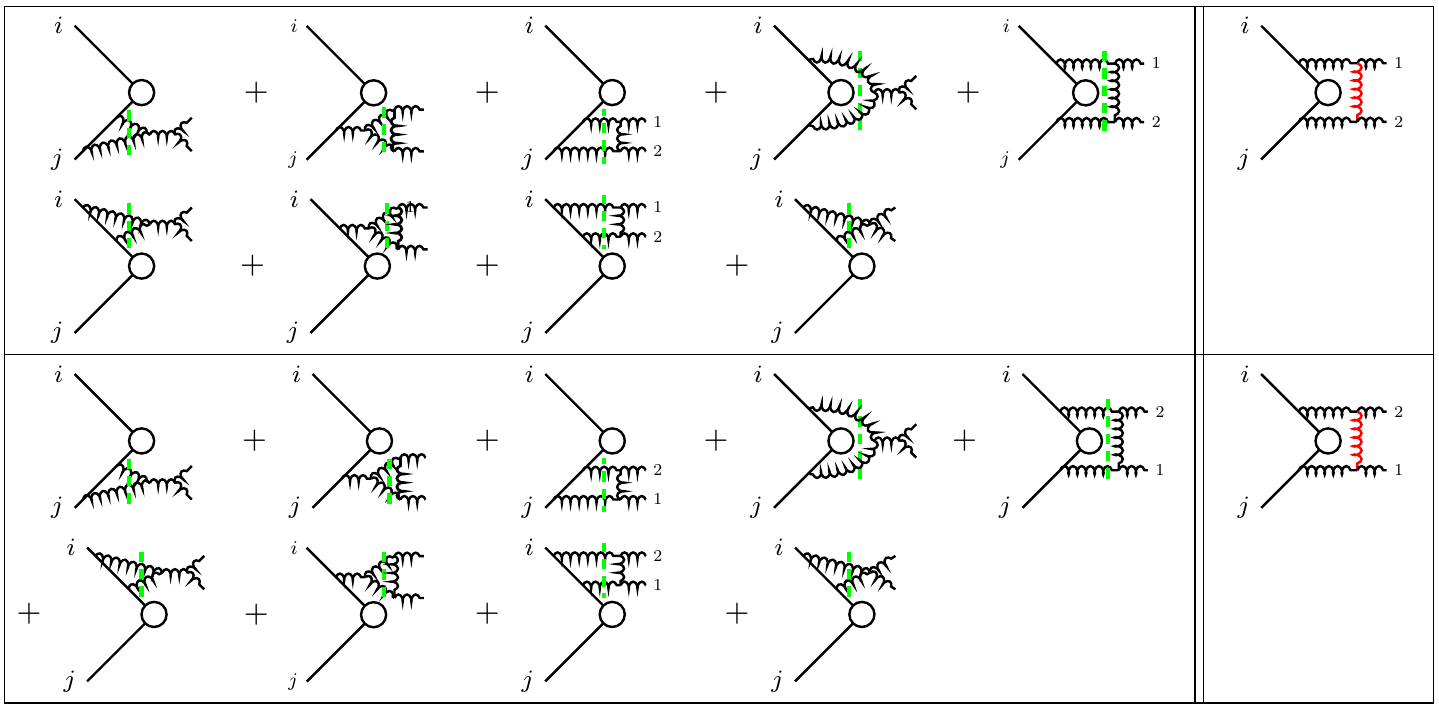} 
\caption{Leading graphs in limits 1--3. Their contributions are
  projected onto the two colour structures in the final column.}
\label{fig:mastergc}
\end{figure}

The leading cuts in limits 1--3 are presented in Fig.~\ref{fig:mastergc} and can be expressed in 
terms of the two colour tensors in Eq.~\eqref{eq:lim1gce}, which are
illustrated in the final column of the figure. There are additional
graphs, other than the ones shown, that involve the four-gluon vertex
but, along with the ghost graphs, these are sub-leading. 
In limit~1 all cuts in this figure are leading except that with a four-gluon vertex. The non-trivial 
way in which these graphs combine to deliver Eq.~\eqref{eq:lim1gce} is
illustrated by considering, as an example, the graphs  that give rise
to the term with Lorentz structure
\begin{align}
-\frac{i\pi}{8\pi^2} \frac{p_j\cdot \vep_1}{p_j\cdot q_1}\frac{q_1\cdot \vep_2}{q_1\cdot q_2}  
\end{align} 
in the first line of Eq.~\eqref{eq:lim1gce}. The first five graphs of
each colour structure are all leading. In the case of the first colour
structure (the top half of Fig.~\ref{fig:mastergc}) we label these $\{G_{1a},G_{1b},G_{1c},G_{1d},G_{1e}\}$.
The first two of these cancel exactly, whilst the others give
\begin{eqnarray}
G_{1c}&=&-\frac{3}{2}\int_{p_j \cdot q_1}^{p_j\cdot q_2}  \frac{\dd
          l_T^2}{l_T^2}  -\frac{3}{2}\int_{0}^{2 q_1 \cdot q_2}  \frac{\dd
          l_T^2}{l_T^2}  ,\\
G_{1d}&=&\frac{3}{2}\int_0^{2q_1\cdot q_2}  \frac{\dd l_T^2}{l_T^2},\\
G_{1e}&=&-\int_0^{2q_1\cdot q_2} \frac{\dd l_T^2}{l_T^2}
+\frac{1}{2}\int^{p_j\cdot q_2}_{p_j\cdot q_1} \frac{\dd l_T^2}{l_T^2} .
\end{eqnarray}
In stark contrast, for the second colour structure the first two
graphs again cancel exactly but the others now give
\begin{eqnarray}
G_{2c}&=&\frac{3}{4}\int_0^{2q_1\cdot q_2}  \frac{\dd
          l_T^2}{l_T^2},\\
G_{2d}&=&-\frac{3}{2}\int_0^{2q_1\cdot q_2}  \frac{\dd l_T^2}{l_T^2},\\
G_{2e}&=&\frac{7}{4}\int_0^{2q_1\cdot q_2} \frac{\dd l_T^2}{l_T^2}
+\int^{p_i\cdot q_2}_{p_i\cdot q_1} \frac{\dd l_T^2}{l_T^2} .
\end{eqnarray}
In both cases, these terms sum up to give the corresponding terms in
Eq.~\eqref{eq:lim1gce}.  

Limit~2 is particularly simple since from all the 
graphs in  Figure  \ref{fig:mastergc}  only graphs $G_{2e}$ and
$G_{2h}$ are leading and they give rise to the two terms
\begin{align}
-\frac{i\pi}{8\pi^2}\frac{1}{2}\frac{p_i\cdot \vep_2}{p_i \cdot q_2}\frac{p_i\cdot \vep_1}{p_i \cdot q_1}
\int_0^{q_{2T}^2}  \frac{\dd l_T^2}{l_T^2}
\end{align}
and
\begin{align}
-\frac{i\pi}{8\pi^2} \left[\frac{1}{2}\frac{p_i\cdot \vep_2}{p_i \cdot q_2}\frac{p_i\cdot \vep_1}{p_i \cdot q_1}-\frac{p_i\cdot \vep_2}{p_i \cdot q_2}\frac{p_j\cdot \vep_1}{p_j \cdot q_1}\right]
\int_0^{q_{2T}^2} \frac{\dd l_T^2}{l_T^2} ~,
 \end{align}
which add up to the corresponding collinear limit of
Eq.~\eqref{eq:lim1gce}. 

Finally we study the leading cuts in limit~3. There 
are leading contributions to the  second colour structure but they
cancel.  The  first colour structure receives leading contributions to the following two Lorentz structures:
\begin{align}
-\frac{i\pi}{8\pi^2}\left\{ \frac{\vep_1^-}{q_1^-} \frac{\vep_2^+}{q_2^+}, \frac{\vep_1^-}{q_1^-} \frac{\vep_2^-}{q_2^-}\right\}~.
\label{eq:auxgc}
\end{align}
Only graph $G_{1e}$ contributes to the first and it gives
\begin{align}
-\int_{0}^{q_{2T}^2}\frac{\dd l_T^2}{l_T^2}.
\end{align}
Graphs $\{G_{1a},G_{1b},G_{1c}, G_{1d}, G_{1e}, G_{1i}\}$ contribute to the second Lorentz structure in  \eqref{eq:auxgc}. The 
contributions of graphs $G_{1a},G_{1b}$ cancel whilst
\begin{eqnarray}
G_{1c}&=&\left[ \frac{3(q_1^{-})^2+3q_1^-q_2^-+2(q_2^-)^2}{4(q_1^-+q_2^-)^2}\right]\int_0^{q_1^+q_2^-}\frac{\dd l_T^2}{l_T^2}-
\frac{1}{2}\int_{p_i\cdot q_1}^{p_i\cdot(q_1 +q_2)}\frac{\dd l_T^2}{l_T^2},\\
G_{1d}&=&- \frac{3q_1^-+2q_2^-}{2(q_1^-+q_2^-)} \int_0^{q_1^+q_2^-}\frac{\dd l_T^2}{l_T^2}+ 
\int_{p_i \cdot q_1}^{p_i\cdot(q_1 +q_2)}\frac{dl_T^2}{l_T^2},\\
G_{1e}&=& \frac{7q_1^-+6q_2^-}{4(q_1^-+q_2^-)}\int_0^{q_{1}^+q_{2}^-}\frac{\dd l_T^2}{l_T^2}
-\frac{1}{2}\int_{p_i \cdot q_1}^{p_i\cdot(q_1 +q_2)}\frac{\dd l_T^2}{l_T^2}
- \int^{p_j\cdot q_1}_{p_j\cdot q_2}\frac{\dd l_T^2}{l_T^2}.
\end{eqnarray}
The sum of these three contributions is
\begin{align}
&\int_0^{q_{2T}^2}\frac{\dd l_T^2}{l_T^2}- \frac{q_1^-q_2^-}{2(q_1^-+q_2^-)^2}
\int_{0}^{q_1^+q_2^-}\frac{\dd l_T^2}{l_T^2}~.
\end{align}
Finally, the four-gluon vertex graph $G_{1i}$ exactly cancels the
second term  of this expression and so the sum of leading graphs in limit~3 reduces to  
\begin{align}
-\frac{i\pi}{8\pi^2}\frac{\vep_1^-}{q_1^-}\left[  \frac{\vep_2^-}{q_2^-} - \frac{\vep_2^+}{q_2^+}\right]
\int_0^{q_{2T}^2}\frac{\dd l_T^2}{l_T^2} \bold{T}_j^{d}if^{dc_2b}if^{bc_1 a}\bold{T}_i^{a}.
\end{align}
This expression is identical to the corresponding collinear limit of Eq.~\eqref{eq:lim1gce}.

It is clear that while the sum of diagrams reproduces
$k_T$ ordering in all three limits, the contributions of individual
diagrams are very different in each region. In particular, the emergence
of $k_T$ ordering in limit~3, of most importance for the understanding
of super-leading logarithms and factorization breaking, involves a very
non-trivial interplay of many different orderings in many different
individual diagrams (in Feynman gauge at least).

\subsection{Physical picture\label{sec:slowphysical}}

In this section, we re-derive the result of the previous section, in a
way that emphasises its physical interpretation. The final result will
be identical to Eq.~(\ref{eq:lim1gce}), but it will be clear that this
represents a $k_T$ ordering in a dipole-like picture: the lower-$k_T$
gluon is emitted by a dipole formed by the higher-$k_T$ gluon and one of
the hard partons, and the $k_T$ of the exchanged Coulomb gluon is
limited by the \emph{local\/} transverse momentum of the softer
gluon, in the frame defined by the dipole of its emission. It is this
local transverse momentum, which is different for different dipoles,
that appears in the logarithms in Eq.~(\ref{eq:lim1gce}).

We consider a general form of the one-loop amplitude to produce two soft
gluons. In particular, we
want to calculate the contribution to the imaginary part of that
amplitude coming from diagrams in which two gluons (or ghosts) are produced at
tree-level and scatter to produce the final-state gluons. 
Suppressing the colour indices, we may write this amplitude as
\begin{align}
  \label{noGhost}
  \hspace*{-2em}
  i\mathcal{M} = \int \frac{\mathrm{d}^{d-2}k_T}{8q_1\ldot
                   q_2(2\pi)^{d-2}\sqrt{1-2k_T^2/(q_1\ldot q_2)}} \times
    \frac12(i\mathcal{A}^{\mu\nu})\Biggl(\sum_{p_1'}\vep^*_{1'\mu;p_1'}\vep_{1'\sigma;p_1'}
    \sum_{p_2'}\vep^*_{2'\nu;p_2'}\vep_{2'\lambda;p_2'}\Biggr)
    (i\mathcal{B}^{\sigma\lambda})~,
  \hspace*{-2em}
\end{align}
where $\mathcal{A}$ is the amplitude to
produce a pair of gluons with momenta $q_1'$ and $q_2'$ and
$\mathcal{B}$ is
the amplitude for that same pair of gluons to scatter into the two final state 
gluons with momenta $q_1$ and $q_2$. The integration variable
$k$ is defined by $q_1' = q_1+k$ and $q_2' = q_2-k$, and the
polarization sums are over physical (on-shell) polarizations of the
cut gluons (labelled by $p_1'$ and $p_2'$).

We will see that our interest is in the region
$k_T^2\ll2q_1\ldot q_2$. In this case, and with physical
polarizations for the gluons, $\mathcal{B}$ is dominated by the
$t$-channel diagram:
\begin{equation}
  \vep_{1'\sigma;p_1'}\vep_{2'\lambda;p_2'}\mathcal{B}^{\sigma\lambda}
  \stackrel{k_T^2\ll2q_1\cdot q_2}\longrightarrow
  g_s^2\mu^{2\epsilon}\,
  \frac{4q_1\ldot q_2}{k_T^2}\,
  \vep_{1';p_1'}\ldot\vep^*_{1;p_1}\,
  \vep_{2';p_2'}\ldot\vep^*_{2;p_2}\,,
\end{equation}
where $p_{1,2}$ label the polarization states of the outgoing
gluons. Therefore we have
\begin{equation}
  \label{finalGeneral}
  \mathcal{M} = \frac{ig_s^2\mu^{2\epsilon}}{8\pi^2}\int \frac{\mathrm{d}^{d-2}k_T}{k_T^2(2\pi)^{d-4}}
    \sum_{p_1'p_2'}
    \Biggl(\mathcal{A}^{\mu\nu}\vep^*_{1'\mu;p_1'}
    \vep^*_{2'\nu;p_2'}\Biggr)
  \vep_{1';p_1'}\ldot\vep^*_{1;p_1}\,
  \vep_{2';p_2'}\ldot\vep^*_{2;p_2}\,.
\end{equation}
When $k_T$ is extremely small, the polarization vector dot-products
become diagonal, i.e.
$\vep_{1';p_1'}\ldot\vep^*_{1;p_1}\to-\delta_{p_1'p_1}$, the momenta
$q'_{1,2}$ become $q_{1,2}$ and there is a complete factorization of the
production, $\mathcal{A}^{\mu\nu}\vep^*_{1\mu;p_1}\vep^*_{2\nu;p_2}$,
from the scattering. However, $k_T^2\ll2q_1\ldot q_2$ is not a
sufficient condition for this and we must evaluate
the polarization vector dot-products more accurately.

\subsection*{Limit~1}

In limit~1 (i.e. the keeping the leading-$\lambda$
terms after the rescaling $q_2\to\lambda q_2$),
$\mathcal{A}^{\mu\nu}$ is a sum of terms each of which has a different
colour factor, and has the form
\begin{equation}
  \label{structure}
  \mathcal{A}^{\mu\nu} \sim \mathcal{A}^\mu
  \left(\frac{p_i^\nu}{p_i\ldot q_2}-\frac{q_1^\nu}{q_1\ldot q_2}\right),
\end{equation}
where $p_i$ is one of the (fast) eikonal-line momenta.
We will find that the region of interest is $k_T^2\ltap2\,p_i\ldot
q_2\,q_1\ldot q_2/p_i\ldot q_1\sim\lambda^2$, so $k_T$ can also be
taken to obey the scaling $k_T\to\lambda k_T$.

In order to evaluate the integral in Eq.~(\ref{finalGeneral}), we
construct explicit representations of the polarization vectors. The
polarization vectors for $q'_{1,2}$ are both perpendicular to both
$q'_1$ and $q'_2$. Obviously the space of such vectors is 2-dimensional
and we can choose basis vectors that span it. To do so, we need an
additional vector, to define the plane of zero azimuth. With an eye on
the structure of Eq.~(\ref{structure}), we use $p_i$ to define this
frame. That is, we take as our polarization vectors
\begin{eqnarray}
  \vep_{1'\mu;\perp} &=& \epsilon_{\mu\nu\lambda\sigma}q'^\nu_1q'^\lambda_2p_i^\sigma
  \,\frac1{\sqrt{2\,p_i\ldot q'_1\,p_i\ldot q'_2\,q'_1\ldot q'_2}}\,, \\
  \vep_{2'\mu;\perp} &=& \epsilon_{\mu\nu\lambda\sigma}q'^\nu_2q'^\lambda_1p_i^\sigma
  \,\frac1{\sqrt{2\,p_i\ldot q'_1\,p_i\ldot q'_2\,q'_1\ldot q'_2}}\,,
\end{eqnarray}
which are perpendicular
to the plane that contains $q'_1$, $q'_2$ and
$p_i$ in the $q'_1\!+\!q'_2$ rest frame, and
\begin{eqnarray}
  \vep_{1'\mu;\parallel} &=& \biggl(q'_1\ldot q'_2\,p_{i\mu}
  -p_i\ldot q'_1\,q'_{2\mu}-p_i\ldot q'_2\,q'_{1\mu}\biggr)
  \,\frac1{\sqrt{2\,p_i\ldot q'_1\,p_i\ldot q'_2\,q'_1\ldot q'_2}}\,, \\
  \vep_{2'\mu;\parallel} &=& \biggl(q'_1\ldot q'_2\,p_{i\mu}
  -p_i\ldot q'_2\,q'_{1\mu}-p_i\ldot q'_1\,q'_{2\mu}\biggr)
  \,\frac1{\sqrt{2\,p_i\ldot q'_1\,p_i\ldot q'_2\,q'_1\ldot q'_2}}\,,
\end{eqnarray}
which are in that
plane. It is worth noting that these statements are
also true in the $p_i\!+\!q'_1$ rest frame: $\perp$ and $\parallel$
states are perpendicular to and in the plane of emission of $q'_2$ in
that frame.
We also define polarization vectors for the gluons with unprimed momenta
in the exactly analogous way. Since the $q_1\!+\!q_2$ rest frame is also
the $q'_1\!+\!q'_2$ rest frame, the two sets of polarization vectors are
related only by rotations.

According to the definition of $k$ above and the fact that two of its
components have been integrated out to put the intermediate gluons
on shell, we have
\begin{eqnarray}
  \label{kinematics1}
  q'_1 &=& (1-\beta)q_1+\beta q_2+k_T \approx
  q_1, \\
  \label{kinematics2}
  q'_2 &=& (1-\beta)q_2+\beta q_1-k_T \approx
  q_2+\frac{k_T^2}{2q_1\ldot q_2}q_1-k_T, \\
  \label{kinematics3}
  \beta &=& \frac12\left(1-\sqrt{1-\frac{2k_T^2}{q_1\ldot q_2}}\right)
  \approx \frac{k_T^2}{2q_1\ldot q_2}\,,
\end{eqnarray}
where, in each case, the first result is exact, while the second is in
the leading-$\lambda$ approximation.

Turning to the polarization vectors in Eq.~(\ref{finalGeneral}), we
first consider the gluon-1 line. We have
\begin{equation}
  \sum_{p_1'}
  \mathcal{A}^\mu\vep^*_{1'\mu;p_1'}\,
  \vep_{1';p_1'}\ldot\vep^*_{1;p_1}\,.
\end{equation}
If $\vep^*_{1;p_1}$ lies in the space spanned by $\vep_{1';p_1'}$, then
this becomes a completeness relation and trivial.
Counting powers of $\lambda$ in all terms, we can show that this is the
case in all of limits~1, 2 and~3, and we can write
\begin{equation}
  \sum_{p_1'}
  \mathcal{A}^\mu\vep^*_{1'\mu;p_1'}\,
  \vep_{1';p_1'}\ldot\vep^*_{1;p_1}
  =-\mathcal{A}^\mu\vep^*_{1\mu;p_1}\,.
\end{equation}
That is, in the leading-$\lambda$ approximation, we can take gluon~1's
momentum and polarization state as being unchanged by the Coulomb
scattering.

Making the same analysis for gluon~2, we find that
the polarization sum is not complete (or rather, the polarization
vector $\vep^*_{2;p_2}$ does not lie in the space spanned by
$\vep_{2';p_2'}$) and therefore we have to evaluate the dot-product
explicitly. Physically, this corresponds to the fact that, although the
scattering is soft, gluon~2 is so much softer than gluon~1 that the
small fraction of gluon~1's momentum that is transferred to gluon~2
does change its momentum and polarization state significantly.

That is, we have to calculate
\begin{equation}
  \mathcal{M} = -\frac{ig_s^2\mu^{2\epsilon}}{8\pi^2}\int \frac{\mathrm{d}^{d-2}k_T}{k_T^2(2\pi)^{d-4}}
  \sum_{p_2'}
  \Biggl(\mathcal{A}^{\mu\nu}
  \vep^*_{1\mu;p_1}\,\vep^*_{2'\nu;p_2'}\Biggr)
  \vep_{2';p_2'}\ldot\vep^*_{2;p_2}.
\end{equation}
To do this, we construct an explicit $k_T$ vector. Since
$q_{1,2}$ and $p_i$ stay fixed during the $k_T$ integration,
$\vep_{1\mu;\perp}$ and $\vep_{1\mu;\parallel}$ are appropriate unit
vectors perpendicular to $q_{1,2}$ and we can write
\begin{equation}
  k_{T\mu} = k_T\sin\phi\,\vep_{1\mu;\perp}
  + k_T\cos\phi\,\vep_{1\mu;\parallel} \,.
\end{equation}
We also note that the dipole form in Eq.~(\ref{structure}) does not
couple to gluons polarized out of the plane and hence the sum over
polarizations $p_2'$ is only over $p_2'=\parallel$.
By explicit calculation, we find that all contributions
integrate to zero, except for diagonal scattering of 
an in-plane polarized gluon scattering to an in-plane polarized gluon
yielding a contribution to the amplitude of
\begin{eqnarray}
  \left(\frac{p_i^\nu}{p_i\cdot q'_2}-\frac{q_1^\nu}{q_1\cdot q_2'}\right)
  \vep^*_{2'\nu;\parallel}\,
  \vep_{2';\parallel}\cdot\vep_{2;\parallel}^* &=& 
  -
  \left(\frac{p_i^\nu}{p_i\cdot q_2}-\frac{q_1^\nu}{q_1\cdot q_2}\right)
  \vep^*_{2\nu;\parallel}\, \nonumber \\ & & \hspace*{-2cm} \times
  \frac{
  1+k_T\cos\phi
  \sqrt{\frac{p_i\cdot q_1}{2\,p_i\cdot q_2\,q_1\cdot q_2}}
  }{1
  +2k_T\cos\phi
  \sqrt{\frac{p_i\cdot q_1}{2\,p_i\cdot q_2\,q_1\cdot q_2}}
  +k_T^2\frac{p_i\cdot q_1}{2\,p_i\cdot q_2\,q_1\cdot q_2}}\,,
\end{eqnarray}
where $\phi=0$ is given by the plane containing $p_i$.
Note that this expression is a function only of $k_T/q_{2T(i1)}$,
where
\begin{equation}
  q_{2T(i1)}^2 = \frac{2\,p_i\cdot q_2\,q_1\cdot q_2}{p_i\cdot q_1}
\end{equation}
is the transverse momentum of gluon~2 in the $p_i\!+\!q_1$ rest frame.

Putting everything together we have
\begin{equation}
  \label{finalintegrand}
  \mathcal{M} =
  \Biggl(\mathcal{A}^{\mu\nu}
  \vep^*_{1\mu;p_1}\,\vep^*_{2\nu;p_2}\Biggr)
  \frac{ig_s^2\mu^{2\epsilon}}{8\pi^2}\int
  \frac{k_T^{d-3}\mathrm{d}k_T\,\mathrm{d}\phi \, \sin^{-2\epsilon}\phi\,\mathrm{d}^{d-4}\Omega}{k_T^2(2\pi)^{d-4}}\,
  \frac{
  1+\frac{k_T}{q_{2T(i1)}}\cos\phi
  }{1
  +2\frac{k_T}{q_{2T(i1)}}\cos\phi
  +\frac{k_T^2}{q_{2T(i1)}^2}}\,.
\end{equation}
The integral can be performed exactly in $d$ dimensions and is
\begin{equation}
  \label{finalresult}
  \mathcal{M} =
  \Biggl(\mathcal{A}^{\mu\nu}
  \vep^*_{1\mu;p_1}\,\vep^*_{2\nu;p_2}\Biggr)
  i\pi\,\frac{\alpha_s}{2\pi}\,
  \frac{\Gamma(1-\epsilon)^2\Gamma(1+\epsilon)}{\Gamma(1-2\epsilon)}
  \left(\frac{4\pi\mu^2}{q_{2T(i1)}^2}\right)^{\epsilon}
  \left(-\frac1\epsilon\right)\,,
\end{equation}
which we have confirmed is in exact agreement with the full calculation from the sum over all
diagrams.

To illustrate the physical structure, it is better to move to four dimensions
where, remarkably, the $\phi$ integral yields an exact $\Theta$-function:
\begin{equation}
  \label{region1final}
  \mathcal{M} =
  \Biggl(\mathcal{A}^{\mu\nu}
  \vep^*_{1\mu;p_1}\,\vep^*_{2\nu;p_2}\Biggr)
  \frac{ig_s^2}{4\pi}\,
  \int_0^{q_{2T(i1)}} \frac{\mathrm{d}k_T}{k_T}\,.
\end{equation}
That is, the $k_T$ of the Coulomb gluon is exactly limited by the
transverse momentum
of the softer of the two gluons it is exchanged between, as measured in
the rest frame of a dipole formed by the harder of the two gluons and
one of the fast partons in the hard process.

In deriving Eq.~(\ref{region1final}), we assumed that
$\mathcal{A}^{\mu\nu}$ has the form of a dipole emission for the
emission of gluon~2 in limit~1. This is, quite generally, the case. In
the particular case of two coloured partons annihilating into a
colourless system, which we
consider in most of this paper, we can explicitly write the final
result as
\bea
&-&\frac{i \pi}{ 8\pi^2 }\left[ \frac{p_j\cdot \varepsilon_1}{p_j\cdot
    q_1}-\frac{p_i\cdot \varepsilon_1}{p_i\cdot q_1}\right]  \left\{
\left[ \frac{p_j\cdot \varepsilon_2}{p_j \cdot q_2}-\frac{q_1\cdot
    \varepsilon_2}{q_1 \cdot q_2}\right]
\int_0^{\frac{2 q_2\cdot q_1\, p_j\cdot q_2}{q_1\cdot p_j}}\frac{\mathrm{d}k_T^2}{k_T^2}\;
\bold{T}_j^{d}if^{dc_2b}if^{bc_1 a}\bold{T}_i^{a} \right.
 \nonumber \\ 
&-& \left. \left[ \frac{p_i\cdot \varepsilon_2}{p_i \cdot q_2}-\frac{q_1\cdot
     \varepsilon_2}{q_1 \cdot q_2}\right]
\int_0^{\frac{2 q_2\cdot q_1\, p_i\cdot q_2}{q_1\cdot p_i}}\frac{\mathrm{d}k_T^2}{k_T^2}\;
\bold{T}_j^{d}if^{dc_1b}if^{bc_2 a}\bold{T}_i^{a}
\right\} \ket{M^{0}}~,
\eea
in agreement with  Eq.~\eqref{eq:lim1gce}.

\subsection*{Limit~2}

Limit~2 is that in which $q_1$ remains fixed and $q_2$ becomes
collinear with one of the hard partons, $p_i$. Much of the previous discussion
applies also
here. In particular, we will take the same forms for the
polarization vectors and momentum exchange, $k$. However, 
some of the power-counting in the
$\lambda\to0$ limit will differ. This is because limit~2 is defined by the scaling
\begin{equation}
  q_2 = \frac{q_2\cdot q_1}{p_i\cdot q_1}p_i +
  \frac{q_{2T}^2}{2q_1\cdot q_2}q_1 + q_{2T}\,;
  \qquad q_{2T}\to\lambda q_{2T}\,;
  \qquad \lambda\to0~.
\end{equation}
Since different components of $q_2$ scale differently with
$\lambda$ we have to be more careful with power counting. For example,
$p_i\sim1$, $q_2\sim1$, but $p_i\cdot q_2\sim\lambda^2$.

In this region, $\mathcal{A}^{\mu\nu}$ factorizes as
\begin{equation}
  \mathcal{A}^{\mu\nu} \sim \mathcal{A}^\mu\mathcal{V}^\nu,
\end{equation}
where the collinear splitting tensor $\mathcal{V}^\nu$ satisfies the
properties
\begin{equation}
  \mathcal{V}^\nu q_{2\nu} = 0
\end{equation}
and
\begin{equation}
  \mathcal{V}^\nu \sim \frac{p_i^\nu}{p_i\cdot q_2}\,
  f\left(\frac{q_2\cdot q_1}{p_i\cdot q_1}\right)+\mathcal{O}(1).
\end{equation}
Note that although $\mathcal{V}^\nu$ scales like $1/\lambda^2$, its
contraction with the polarization vector $\vep^*_{2\nu;p_2}$ is
proportional to $p_i\cdot\vep^*_{2;p_2}$ so in the collinear limit
$p_i\sim q_2$, $q_2\cdot\vep^*_{2;p_2}=0$ implies
$p_i\cdot\vep^*_{2;p_2}\sim\lambda$ and hence the physical amplitude
scales like
\begin{equation}
  \label{collinearscaling}
  \mathcal{V}^\nu \vep^*_{2\nu;p_2} \sim \frac1{\lambda}\,.
\end{equation}
Another difference relative to the case of limit~1 is that $\mathcal{V}^\nu$
couples to both polarizations of gluon, not only in-plane polarization.

Having made these preliminary remarks, most of the calculation is the
same as before. In particular, the first equality in
each of Eqs.~(\ref{kinematics1}--\ref{kinematics3}) is
unchanged. However, this time $\beta\sim\lambda^2$, so the scattering is
extremely soft. This means that the change in direction of $q_1$, and
hence of its polarization vector, is even smaller than it is in limit~1 and
hence we can continue to assume that its polarization sum is
complete. On the other hand, we find that
although the change in direction of gluon~2 is much smaller than in
limit~1, it is equally important, because gluon~2 is much closer to the
collinear direction and hence a small change in direction changes the
amplitude significantly.  The final result for
$\vep_{2';\parallel}\cdot\vep_{2;\parallel}^*$ is unchanged. Moreover, the property
Eq.~(\ref{collinearscaling}) implies
\begin{equation}
  \mathcal{V}^\nu(q_2')\vep_{2'\nu;p_2}^* =
  \mathcal{V}^\nu(q_2)\vep_{2\nu;p_2}^*\sqrt{\frac{p_i\cdot q_2}{p_i\cdot q_2'}}
  \,.
\end{equation}
Finally, therefore, the result for an in-plane polarized gluon
scattering to an in-plane polarized gluon is identical to the one in
limit~1.

We also have to calculate the production of an out-of-plane polarized
gluon scattering to either an in-plane or out-of-plane polarized
gluon. It is a few lines of calculation to show that the off-diagonal
scattering again integrates to zero, and the result for
$\vep_{2';\perp}\cdot\vep_{2;\perp}^*$ is identical to
$\vep_{2';\parallel}\cdot\vep_{2;\parallel}^*$.
The final result is therefore identical to that in limit~1, i.e.
Eq.~(\ref{finalintegrand}).

\subsection*{Limit~3}
Limit~3 is relevant for the super-leading logarithms discovered in
\cite{Forshaw:2006fk} and in this case
gluon~1 is collinear with one of the hard partons, $p_i$:
\begin{equation}
  q_1 \to \frac{q_1\cdot q_2}{\lambda\, p_i\cdot q_2}p_i +
  \frac{\lambda\, q_{1T}^2}{2q_1\cdot q_2}q_2 + q_{1T}\, ,
\end{equation}
and gluon~2 is soft relative to gluon~1, i.e. $  q_2\to\lambda q_2$.
Note that the $q_2$ direction is used as the reference for the collinear
limit of $q_1$, and that $\lambda$ controls both the collinear limit of
gluon~1 and the soft limit of gluon~2. In this limit, $\mathcal{A}^{\mu\nu}$ factorizes as
\begin{equation}
  \mathcal{A}^{\mu\nu} \sim \mathcal{V}^\mu\mathcal{A}^\nu.
\end{equation}
The collinear splitting tensor $\mathcal{V}^\mu$ scales as
$\sim1/\lambda$, but $\mathcal{V}\cdot\vep_1^*\sim1$, while
$\mathcal{A}^\nu\sim\mathcal{A}\cdot\vep_2^*\sim1/\lambda$.

The amplitude $\mathcal{A}^\nu$ contains contributions to the emission
of $q_2$ from the $p_i\!-\!q_1$ dipole and also from the $p_j\!-\!q_1$ dipole.
Since $q_1$ is becoming collinear with $p_i$ and $q_2$ is
being emitted at a large angle to them, emission from the $p_i\!-\!q_1$
dipole is suppressed by a power of $\lambda$, since
it corresponds to emission far outside the angular region of the dipole.
So only emission of~$q_2$ from the $p_j\!-\!q_1$ dipole is leading.

With this in mind, we use $p_j$ to fix the $\phi=0$ plane rather
than~$p_i$. With this exception, the definitions of the kinematics and
polarization vectors, and most of the rest of the calculation, are the
same as in limit~1. We again find that the relevant region is $k_T\ltap
q_{2T}$, where $q_{2T}$ is defined in the $p_j\!-\!q_1$ dipole frame,
and hence $k_T\sim\lambda$, giving $\beta\sim\lambda^2$.

Considering the gluon-1 line, even though gluon~1 is collinear with $p_i$, its
shift in direction due to the Coulomb exchange with the even softer
gluon~2 is so small that we can once again ignore it. For the gluon-2 line, the expressions for the amplitude
and polarization dot-product
are the same, but with $p_i$ replaced by $p_j$.
Thus, finally, the result is exactly the same as in limit~1.

\section{Conclusions}

Attention has been focussed over recent years on the role of Coulomb
gluon exchange in partonic scattering, in part spurred on by
the discovery of super-leading
logarithmic terms in gaps-between-jets and the fact that they give rise
to violations of coherence and collinear factorization. Previous analyses have been based on the
colour evolution picture, in which it is assumed that the evolution is
ordered in transverse momentum of exchanged and emitted gluons. In this
paper we have made substantial progress in confirming the validity of
this assumption. We did this by
making a full Feynman diagrammatic calculation of the one-loop
correction to a colour annihilation process accompanied by the emission
of up to two gluons. Although the result for individual diagrams is
complicated and different diagrams clearly have different ordering
conditions, the result for the physical process, i.e.~the sum of all
diagrams, is very simple: the exchange of the Coulomb gluon is ordered
in transverse momentum with respect to the transverse momenta of the
emitted gluons.


Although we have focussed on one-loop corrections to processes with
incoming partons only, and up to two emitted gluons only, most of our
calculation can be generalised rather easily to processes with outgoing
partons and any number of emitted gluons, and we will do this in
forthcoming work.

Our calculation has also provided further insight into the structure of
Coulomb gluon corrections. Specifically, we have seen that the full
emission and exchange process can be separated gauge-invariantly into
distinct physical processes (Figs.~\ref{fig:two} and~\ref{fig:gc}). Each
process corresponds to Coulomb exchange in the distant past
or future, with gluon emission from the hard process or any of the
exchange processes. Perhaps this offers hope of a deeper
understanding of the role of Coulomb gluons and a generalization of our
calculation to an arbitrary number of loops.

\acknowledgments
We should like to thank to George Sterman, Zoltan Nagy, Mrinal Dasgupta and Alessandro Fregoso for many enjoyable and helpful discussions. 
This work is partially supported by the Lancaster-Manchester-Sheffield
Consortium for Fundamental Physics under STFC grants ST/J000418/1 and
ST/L000520/1, by the FP7 Marie Curie Initial Training Network MCnetITN
under European Union grant PITN-GA-2012-315877 and by the
Mexican National Council of Science and Technology (CONACyT).

\bibliography{bordering}
\bibliographystyle{JHEP.bst}

\end{document}